\documentclass[12pt]{article}
\usepackage{amsmath}
\usepackage{graphicx}
\usepackage[utf8]{inputenc}

\usepackage[hyphens]{url} 
\usepackage{hyperref}
\providecommand{\tightlist}{%
  \setlength{\itemsep}{0pt}\setlength{\parskip}{0pt}}

\usepackage{booktabs}
\usepackage{longtable}
\usepackage{amsmath}
\usepackage{amssymb}
\usepackage{xcolor}
\usepackage{bm}
\usepackage{authblk}

\begin{document}

\title{Using tours to visually investigate properties of new projection pursuit
indexes with application to problems in physics }

\author[1,2]{Ursula Laa}
\author[2]{Dianne Cook}

\affil[1]{School of Physics and Astronomy, Monash University}
\affil[2]{Department of Econometrics and Business Statistics, Monash University}

\maketitle

\begin{abstract}
Projection pursuit is used to find interesting low-dimensional
projections of high-dimensional data by optimizing an index over all
possible projections. Most indexes have been developed to detect
departure from known distributions, such as normality, or to find
separations between known groups. Here, we are interested in finding
projections revealing potentially complex bivariate patterns, using new
indexes constructed from scagnostics and a maximum information
coefficient, with a purpose to detect unusual relationships between
model parameters describing physics phenomena. The performance of these
indexes is examined with respect to ideal behaviour, using simulated
data, and then applied to problems from gravitational wave astronomy.
The implementation builds upon the projection pursuit tools available in
the R package, tourr, with indexes constructed from code in the R
packages, binostics, minerva and mbgraphic.
\\

\end{abstract}

\def\spacingset#1{\renewcommand{\baselinestretch}%
{#1}\small\normalsize} \spacingset{1}

\hypertarget{introduction}{%
\section{Introduction}\label{introduction}}

The term ``projection pursuit'' (PP) was coined by Friedman and Tukey
(1974) to describe a procedure for searching high (say
\(p-\))dimensional data for ``interesting'' low-dimensional projections
(\(d=1\) or \(2\) usually). The procedure, originally suggested by
Kruskal (1969), involves defining a criterion function, or index, that
measures the ``interestingness'' of each \(d\)-dimensional projection of
\(p\)-dimensional data. This criterion function is optimized over the
space of all \(d\)-dimensional projections of \(p\)-space, searching for
both global and local maxima. It is hoped that the resulting solutions
reveal low-dimensional structure in the data not found by methods such
as principal component analysis. Projection pursuit is primarily used
for visualization, with the projected data always reported as plots.

A large number of projection pursuit indexes have been developed,
primarily based on departure from normality, which includes clusters,
outliers and skewness, and also for finding separations between known
groups (e.g.~Friedman (1987), Hall (1989) Cook, Buja, and Cabrera
(1992), Naito (1997), Lee et al. (2005), Ahn, Hofmann, and Cook (2003),
Hou and Wentzell (2014), Jones and Sibson (1987), Rodriguez-Martinez et
al. (2010), Pan, Fung, and Fang (2000), Ferraty et al. (2013), Loperfido
(2018)). Less work has been done on indexes to find nonlinear dependence
between variables, focused on \(d=2\), which motivates this research.

The driving application is from physics, to aid the interpretation of
model fits on experimental results. A physical model can be considered
to be a set of \(p\) free parameters, that cannot be measured directly
and are determined by fitting a set of \(q~ (p<q)\) experimental
observations, for which predictions can be made once the \(p\)
parameters are estimated. (Note here, that while we may have analytic
expressions for the predictions, this is not always the case and we
often have to rely on numerical computation.) Different sets of model
parameters (\(n\)) found to be compatible with the experimental results
within a selected level of confidence yield the data to be examined
using projection pursuit. A single prediction can be a complicated
function of all of the free parameters, and typically
\(q \in [100,1000]\) and \(p \sim 10\). Current practice is to examine
pairs of parameters, or combinations produced by intuition or prior
knowledge. This begs the question, whether important nonlinear
associations are missed because they are hidden in linear combinations
of more than two variables.

PP can be combined with other dimension reduction methods when \(p\) is
very high. For example, it can be beneficial to first do principal
component analysis prior to PP, especially to remove linear dependencies
before searching for other types of association. This is the approach
used in Cook, Laa, and Valencia (2018), which explores a 56-dimensional
parameter space, by first reducing the number of dimensions to the first
six principal components, before applying projection pursuit. PCA was
appropriate for this problem because reducing to principal component
space removed the linear dependencies while preserving the nonlinear
relationships that were interesting to discover. Some projection pursuit
indexes do incorporate penalty terms to automate removing noise
dimensions. It can also be important to have an efficient PP optimizer,
particularly when working with high dimensions, because the search space
increases exponentially with dimension.

To find appropriate projections pursuit indexes for detecting nonlinear
dependencies, the literature on variable selection was a starting point.
With high-dimensional data, even plotting all pairs of variables can
lead to too many plots, which is what ``scagnostics'' (Wilkinson, Anand,
and Grossman (2005), Wilkinson and Wills (2008)) were developed to
address by providing metrics from which to select the most interesting
variable pairs. There are eight scagnostics, of which three (``convex'',
``skinny'' and ``stringy'') are used here. The question is whether these
can be adapted into projection pursuit indexes, to search for unusual
features in two-dimensional projections of high-dimensional data. Recent
PhD research by Grimm (2016) explored the behavior of scagnostics for
selecting variables, and proposed two more that have nicer properties,
based on smoothing splines and distance correlation. In addition, two
more indexes for measuring dependence have been proposed in the machine
learning literature, based on information criteria, maximal and total
information coefficient (MIC and TIC) (Reshef et al. 2011), with
computationally more efficient versions (MIC\_e, TIC\_e) (Reshef et al.
2016). These are related to original 1D projection pursuit indexes based
on entropy (e.g.~Huber (1985), Jones and Sibson (1987)). This provides
seven current indexes for measuring dependence between two variables,
and each is available in an R (R Core Team 2018) package: binostics
(Hofmann et al. 2019), mbgraphic (Grimm 2017) and minerva (Albanese et
al. 2012).

PP index behavior can be understood and investigated more substantially
when combined with a tour. A tour (Asimov (1985), Buja et al. (2005))
displays a smooth sequence of low dimensional projections from high
dimensions to explore multivariate data for structure such as clusters,
outliers, and nonlinear relationships. Cook et al. (1995) provided an
approach combining the tour algorithm with PP, to interactively both
search for interesting projections, and examine the behavior of the
indexes. The projection pursuit guided tour is available in the R
package, tourr (Wickham et al. 2011), and provides optimization
routines, and visualization.

This paper is structured as follows. Section \ref{sec:construct}
discusses index construction, and how they can be used in the guided
tour. Section \ref{sec:investigate} investigates the behavior of the
indexes, explored primarily using tour methods. The new guided tour with
these indexes is applied to two examples from gravitational wave
astronomy (Section \ref{sec:phys}). The latter two parts are connected
in that the application of the new indexes to these problems is the main
motivation for the paper, and the simulation study, in the first part,
was conducted to better understand the behavior of the indexes in
general. The techniques in Section \ref{sec:investigate} define
procedures that will be generally useful for researchers developing new
projection pursuit indexes to visually assess their behavior. Visual
methods to diagnose the index behavior is important because PP is
primarily used for visualization. The paper finishes with a discussion
about the limitations of this work, and the potential future directions.

\hypertarget{projection-pursuit-index-construction-and-optimization}{%
\section{Projection pursuit index construction and
optimization}\label{projection-pursuit-index-construction-and-optimization}}

\label{sec:construct}

A projection pursuit index (PPI) is a scalar function \(f\) defined on
an \(d\)-dimensional data set, computed by taking a \(d\)-dimensional
projection of an \(n\times p\) data matrix. Typically the definition is
such that larger values of \(f\) indicate a more interesting
distribution of observations, and therefore maximizing \(f\) over all
possible \(d\) dimensional projections of a data set with \(p>d\)
variables will find the most interesting projections. This section
describes the seven indexes that are to be used to explore bivariate
association. Some data pre-processing, including standardization, is
advisable, prior to optimizing the PPI.

\hypertarget{scaling-and-standardization}{%
\subsection{Scaling and
standardization}\label{scaling-and-standardization}}

Making a plot always involves some choice of scaling. When a scatterplot
is made, effectively, albeit under the hood, the data is scaled into a
range of \([a, b]\) (often \(a=0, b=1\)) on both axes to print it on a
page or display in a graphics device window. The range deliminates page
space within which to draw. The upshot is that the original data scale
is standardized to the range and aspect ratio on the display space. It
may be that the original range of one variable is \([1, 1000]\) and the
other is \([1, 1.6]\) but the display linearly warps this to \([0, 1]\)
and \([0, 1]\), say, giving both variables equal visual weight.

With high dimensional data, and particularly projections, it is also
necessary to re-scale the original range, and it is important to pay
attention to what is conventional, or possible, and the effects. The
PPIs also may require specific scaling for them to be effectively
computed. Both of these are addressed here. The common pre-processing
include:

\begin{itemize}
\tightlist
\item
  Standardizing each variable, to have mean 0 and variance 1, so that
  individual variable scales do not affect the result. Different
  variable scales are examinable without resorting to projection
  pursuit, so can be handled prior to searching through high dimensions.
\item
  Sphering the high-dimensional data is often done to remove linear
  dependence. This is typically done using principal component analysis,
  and using the principal components as the variables passed to PP. If
  linear dependence is the only structure PP is not needed, and thus
  this is removed before PP so that the PPIs are not distracted by
  simple structure.
\item
  Transform single variables to reduce skewness. It is marginal
  structure, visible in a single variable, which doesn't need a
  multivariate technique to reveal. Skewed distributions will
  inadvertently affect the PPIs, distracting the search for dependence.
\item
  Remove outliers, which may be an iterative process, to discover,
  identify and delete. Extreme values will likely affect PPI
  performance. Outliers can be examined on a case by case basis later.
\item
  Possibly remove noise dimensions, which is also likely to be an
  iterative process. Directions where the distribution is purely noise
  make optimization of a PPI more difficult. If a variable is suspected
  to have little structure and relationship with other variables,
  conducting PP on the subset of variables without them may improve the
  efficiency of the search.
\item
  Centering and scaling of the projected data, can be helpful visually.
  If the data has a small amount of non-normal distribution in some
  directions, the projected data can appear to wander around the plot
  window during a tour. It doesn't matter what the center of the
  projected data is, so centering removes a wandering scatterplot. Less
  commonly, it may be useful to scale the projected data to standard
  values, which would be done to remove any linear dependence remaining
  in the data.
\end{itemize}

\begin{table}[htp]
\caption{Summary of notation}
\centering
\begin{tabular}{|l|p{8.2cm}|} \hline
$p$ & be the number of variables in the data. For physics models the observable space is higher dimension, say $q$, and the data examined is the fitted model space, typically less than 10. \\
$d (=2) $ & projection dimension. For studying physics models, typically $d=2$ and this is the focus for the index definition. \\
$n$ & Number of observations, which for the physics models is the number of fitted models being examined and compared. \\
&\\
$\bm{X}_{n\times p} = (\bm{X}_1, ..., \bm{X}_p)$ & $n\times p$-dimensional data matrix, where variables $\bm{X}_j$ may be scaled or standardized, and $\bm{X}$ may be sphered\\
$\bm{Y}_{n\times 2} = (\bm{Y}_1, ..., \bm{Y}_2)$ & projected data matrix. where $Y_j = (\alpha_1 X_1, ..., \alpha_p X_p)$ \\
$\bm{F}_{p\times 2} = (\bm{\alpha}_1, ..., \bm{\alpha}_2)$ & orthonomal projection matrix\\ 
&\\
$H$ & Convex hull \\
$A$ & Alpha hull \\\hline
\end{tabular}
\label{notation}
\end{table}

\hypertarget{new-projection-pursuit-index-functions}{%
\subsection{New projection pursuit index
functions}\label{new-projection-pursuit-index-functions}}

\label{sec:indexDef}

Table \ref{notation} summarizes the notation used for this section. Here
we give an overview of the functions that are converted into projection
pursuit indexes. Full details of the functions can be found in the
original sources.

\begin{itemize}
\tightlist
\item
  \textbf{scagnostics}: The first step to computing the scagnostics is
  that the bivariate data is binned, and scaled between {[}0,1{]} for
  calculations. The convex (Eddy 1977) and alpha hulls (Edelsbrunner,
  Kirkpatrick, and Seidel 1983), and the minimal spanning tree (MST)
  (Kruskal 1956), are computed.

  \begin{itemize}
  \tightlist
  \item
    \emph{convex}: The ratio of the area of alpha to convex hull,
    \(I_{\bm{F}, convex}= \frac{area(A(Y))}{area(H(Y))}\). This is the
    only measure where interesting projections will take low values,
    with a maximum of 1 if both areas are the same. Thus
    \(1-c_{convex}\) is used.
  \item
    \emph{skinny}: The ratio of the perimeter to the area of the alpha
    hull,
    \(I_{\bm{F}, skinny} = 1 - \frac{\sqrt{4\pi area(A)}}{perimeter(A)}\),
    where the normalization is chosen such that
    \(I_{\bm{F}, skinny} = 0\) for a full circle. Values close to 1
    indicate a skinny polygon.
  \item
    \emph{stringy}: Based on the MST,
    \(I_{\bm{F}, stringy} = \frac{diameter(MST)}{length(MST)}\) where
    the diameter is the longest connected path, and the length is the
    total length (sum of all edges). If the MST contains no branches
    \(I_{\bm{F}, skinny} = 1\).
  \end{itemize}
\item
  \textbf{association}: The index functions are available in Grimm
  (2017), and are defined to range in {[}0,1{]}. Both functions in the
  mbgraphic package can bin the data before computing the index, for
  computational performance.

  \begin{itemize}
  \tightlist
  \item
    \emph{dcor2D}: This function is based on distance correlation
    (Székely, Rizzo, and Bakirov 2007), which is designed to find both
    linear and nonlinear dependencies between variables. It involves
    computing the distances between pairs of observations, conducting an
    analysis of variance type breakdown of the distances relative to
    each variable, and the result is then passed to the usual co
    variance and hence correlation formula. The function wdcor, in the R
    package, extracat (Pilhöfer and Unwin 2013), computes the statistic,
    and the mbgraphic package utilises this function.
  \item
    \emph{splines2D}: Measures nonlinear dependence by fitting a spline
    model (Wahba 1990) of \(\bm{Y}_2\) on \(\bm{Y}_1\) and also
    \(\bm{Y}_1\) on \(\bm{Y}_2\), using the \texttt{gam} function in the
    R package, mgcv (Wood et al. 2016). The index compares the variance
    of the residuals: \begin{equation}
      I_{\bm{F}, splines2d} = max(1- \frac{Var(res_{\bm{Y}_1\sim \bm{Y}_2})}{Var(\bm{Y}_1)}, 1-\frac{Var(res_{\bm{Y}_2\sim {\bm{Y}_1}})}{Var(\bm{Y}_2)}),
      \end{equation} which takes large values if functional dependence
    is strong.
  \end{itemize}
\item
  \textbf{information}: The index functions (Reshef et al. 2011)
  nonparametricly measure nonlinear association by computing the mutual
  information, \begin{equation}
  \bm{I} = \sum_{by_1}\sum_{by_2} p(by_1, by_2) log(p(by_1, by_2)/(p(by_1)p(by_2))),
  \end{equation} where \(by_1, by_2\) are binned values of the projected
  data, and \(p(by_1, by_2)\)is the relative bin count in each cell, and
  \(p(by_1), p(by_2)\) are the row and column relative counts, on a
  range of bin resolutions of the data. It is strictly a 2D measure. For
  a fixed binning, e.g.~\(2\times 2\) or \(10\times 4\), the optimal
  binning is found by maximizing \(I\). The values of I range between
  \([0,1]\) because they are normalized across bins by dividing by
  \(log(min(\# bins_{y_1}, \# bins_{y_2}))\).

  \begin{itemize}
  \tightlist
  \item
    \emph{Maximum Information Coefficient (MIC)}: uses the maximum
    normalized \(I\) across all bin resolutions.
  \item
    \emph{Total Information Coefficient (TIC)}: sums the normalized
    \(I\) for all bin resolutions computed. This creates a problem of
    scaling - there is no upper limit, although it is related to number
    of bins, and number of bin resolutions used. In the work below we
    have made empirical estimates of the maximum and scaled the TIC
    index using this to get it in the range \([0,1]\). This index should
    be more stable than MIC.
  \end{itemize}
\end{itemize}

A comparison between these indexes for the \textbf{purpose of variable
selection}, but not projection pursuit, was discussed in Grimm (2016).
It is only available in German, we we summarize the main findings here.
The scagnostics measures are flexible, and calculating the full set of
measures provides useful guidance in variable selection. However, they
are found to be highly sensitive to outlying points and sample size (as
a consequence of the binning). Both splines2D and dcor2D are found to be
robust in this respect, but splines2D is limited to functional
dependence, while dcor2D is found to take large values only in scenarios
with large linear correlation. The mutual information based index
functions (MIC, TIC) are found to be flexible, but are sensitive to the
sample size and often take relatively large values even when no
association is present. A brief comparison of MIC and dcor2D was also
provided in Simon and Tibshirani (2014).

In addition to the seven indexes described above, we will also include
the holes index available in the \texttt{tourr} package, see (Cook,
Buja, and Cabrera (1993), Cook and Swayne (2007)). This serves as a
benchmark, demonstrating some desired behavior. The index takes maximum
values for a central hole in the distribution of the projected data.

\hypertarget{optimization}{%
\subsection{Optimization}\label{optimization}}

Given a PPI we are confronted with the task of finding the maximum over
all possible \(d\) dimensional projections. One challenge is to avoid
getting trapped in local maxima that are only a result of sampling
fluctuations or a consequence of a noisy index function. Posse (1995a)
discusses the optimization, in particular that for most index functions
and optimizers results are too local, largely dependent on starting
point. Friedman (1987) suggested a two-step procedure: the first step is
using a large step size to find the approximate global maximum while
stepping over pseudomaxima. A second step is then starting from the
projection corresponding to the approximate maximum and employing a
gradient directed optimization for the identification of the maximum.
For exploring high-dimensional data, it can be interesting to observe
local maxima as well as a global maximum, and thus a hybrid algorithm
that still allows lingering but not being trapped by local maxima is
ideal. In addition, being able to visually monitor the optimization and
see the optimal projection in the context of neighboring projections is
useful. This is provided by combining projection pursuit with the grand
tour (Cook et al. 1995). The properties of a suitable optimization
algorithm include monotonicity of the index value, a variable step-size
to avoid overshooting and to increase the chance of reaching the nearest
maximum, and a stopping criterion allowing to move out of a local
maximum and into a new search region (Wickham et al. 2011). A possible
algorithm is inspired by simulated annealing and has been described in
Lee et al. (2005), this has been implemented in the
\texttt{search\_better} and \texttt{search\_better\_random} search
functions in the tourr package. The tourr package also provides the
\texttt{search\_geodesic} function, which first selects a promising
direction by comparing index values between a selected number of small
random steps, and then optimizes the function over the line along the
geodesic in that direction considering projections up to \(\pi/4\) away
from the current view.

\hypertarget{investigation-of-indexes}{%
\section{Investigation of indexes}\label{investigation-of-indexes}}

\label{sec:investigate}

A useful projection pursuit index needs to have several properties. This
has been discussed in several seminal papers, e.g.~Diaconis and Freedman
(1984), Huber (1985), Jones and Sibson (1987), Posse (1995a), Hall
(1989). The PPI should be minimized by the normal distribution, because
this is not interesting from a data exploration perspective. If all
projections are normally distributed, good modelling tools already
exist. A PPI should be approximately affine invariant, regardless how
the projection is rotated the index value should be the same, and the
scale of each variable shouldn't affect the index value. Interestingly
the original index proposed by Friedman and Tukey (1974) was not
rotationally invariant. A consistent index means that small
perturbations to the sample do not dramatically change the index value.
This is particularly important to making optimization feasible, small
angles between projections correspond to small perturbations of the
sample, and thus should be small changes to index value. Posse (1995a)
suggests that indexes should be resistant to features in a tail of the
distribution, but this is debatable because one departure from normality
that is interesting to detect are anomalous observations. Some PPI are
designed precisely for these reasons. Lastly, because we need to compute
the PPI over many projections, it needs to be fast to compute. These
form the basis of the criteria upon which the scagnostic indexes, and
the several alternative indexes are examined, as explained below.

\begin{itemize}
\item
  \textbf{smoothness}: This is the consistency property mentioned above.
  The index function values are examined over interpolated tour paths,
  that is, the value is plotted against time, where time indexes the
  sequence of projections produced by the tour path. The signature of a
  good PPI is that the plotted function is relatively smooth. The
  interpolation path corresponds to small angle changes between
  projections, so the value should be very similar.
\item
  \textbf{squintability}: Tukey and Tukey (1981) introduced the idea of
  squint angle to indicate resolution of structure in the data. Fine
  structure like the parallel planes in the infamous RANDU data
  (Marsaglia 1968) has a small squint angle because you have to be very
  close to the optimal projection plane to be able to see the structure.
  Structures with small squint angle are difficult to find, because the
  optimization algorithm needs to get very close to begin hill-climbing
  to the optimum. The analyst doesn't have control over the data
  structure, but does have control over the PPI. Squintability is about
  the shape of the PPI over all projections. It should have smooth low
  values for noise projections and a clearly larger value, ideally with
  a big squint angle, for structured projections. The optimizer should
  be able to clearly see the optimal projections as different from
  noise. To examine squintability, the PPI values are examined on
  interpolated tour paths between a noise projection and a distant
  structured projection.
\item
  \textbf{flexibility}: An analyst can have a toolbox of indices that
  may cover the range of fine and broad structure, which underlies the
  scagnostics suite. Early indexes, based on density estimation could be
  programmed to detect fine or large structure by varying the binwidth.
  This is examined by using a range of structure in the simulated data
  examples.
\item
  \textbf{rotation invariance}: The orientation of structure within a
  projection plane should not change the index value. This is especially
  important when using the projection pursuit guided tour, because the
  tour path is defined between planes, along a geodesic path, not bases
  within planes. If a particular orientation is more optimal, this will
  get lost as the projection shown pays no attention to orientation.
  Buja et al. (2005) describes alternative interpolation paths based on
  Givens and Householder rotations which progress from basis to basis.
  It may be possible to ignore rotation invariance with these
  interpolations but there isn't a current implementation, primarily
  because the within-plane spin that is generated is distracting from a
  visualization perspective. Rotation invariance is checked for the
  proposed PPIs by rotating the structured projection, within the plane.
\item
  \textbf{speed}: Being fast to compute allows the index to be used in
  real-time in a guided tour, where the optimization can be watched.
  When the computations are shifted off-line, to watch in replay,
  computation times matter less. This is checked by comparing times for
  benchmark scenarios with varying sample size.
\end{itemize}

\hypertarget{simulation-study-setup}{%
\subsection{Simulation study setup}\label{simulation-study-setup}}

\hypertarget{data-construction}{%
\subsubsection{Data construction}\label{data-construction}}

\label{sec:dataOv}

Three families of data simulations are used for examining the behavior
of the index functions. Each generates structure in two variables, with
the remaining variables containing various types of noise. This is a
very simple construction, because there is no need for projection
pursuit to find the structure, one could simply use the PPIs on pairs of
variables. However, it serves the purpose to also evaluate the PPIs. The
three data families are explained below. In each set, \(n\) is used for
the number of points, \(p\) is the number of dimensions, and \(d=2\) is
the projection dimension. The three structures were selected to cover
both functional and non-functional dependence, different types of
nuisance distributions and different structure size and squintability
properties.

\begin{itemize}
\tightlist
\item
  \textbf{pipe}: nuisance directions are generated by sampling
  independently from a uniform distribution between \([-1,1]\), and the
  circle is generated by sampling randomly on a 2D circle, and adding a
  small radial noise. The circle should be easy to see by some indices
  because it is large structure, but the nonlinearity creates a
  complication.
\item
  \textbf{sine}: nuisance directions are generated by sampling
  independently from a standard normal distribution, and the sine curve
  is generated by \(x_p = \sin(x_{p-1}) + \mathrm{jittering}\). The sine
  is a medium nonlinear structure, which should be visible to multiple
  indices.
\item
  \textbf{spiral}: nuisance directions are generated by sampling
  independently from a normal distribution, and the structure directions
  are sampled from an Archimedean spiral, i.e.~\(r = a + b \theta\),
  with \(a=b=0.1\) and we sample angles \(\theta\) from a normal
  distribution with mean 0 and variance \(2\pi\), giving a spiral with
  higher densities at lower radii. The absolute value of \(\theta\)
  fixes the direction of the spiral shape. This is fine structure which
  is only visible close to the optimal projection.
\end{itemize}

\begin{figure}

{\centering \includegraphics[width=0.8\textwidth]{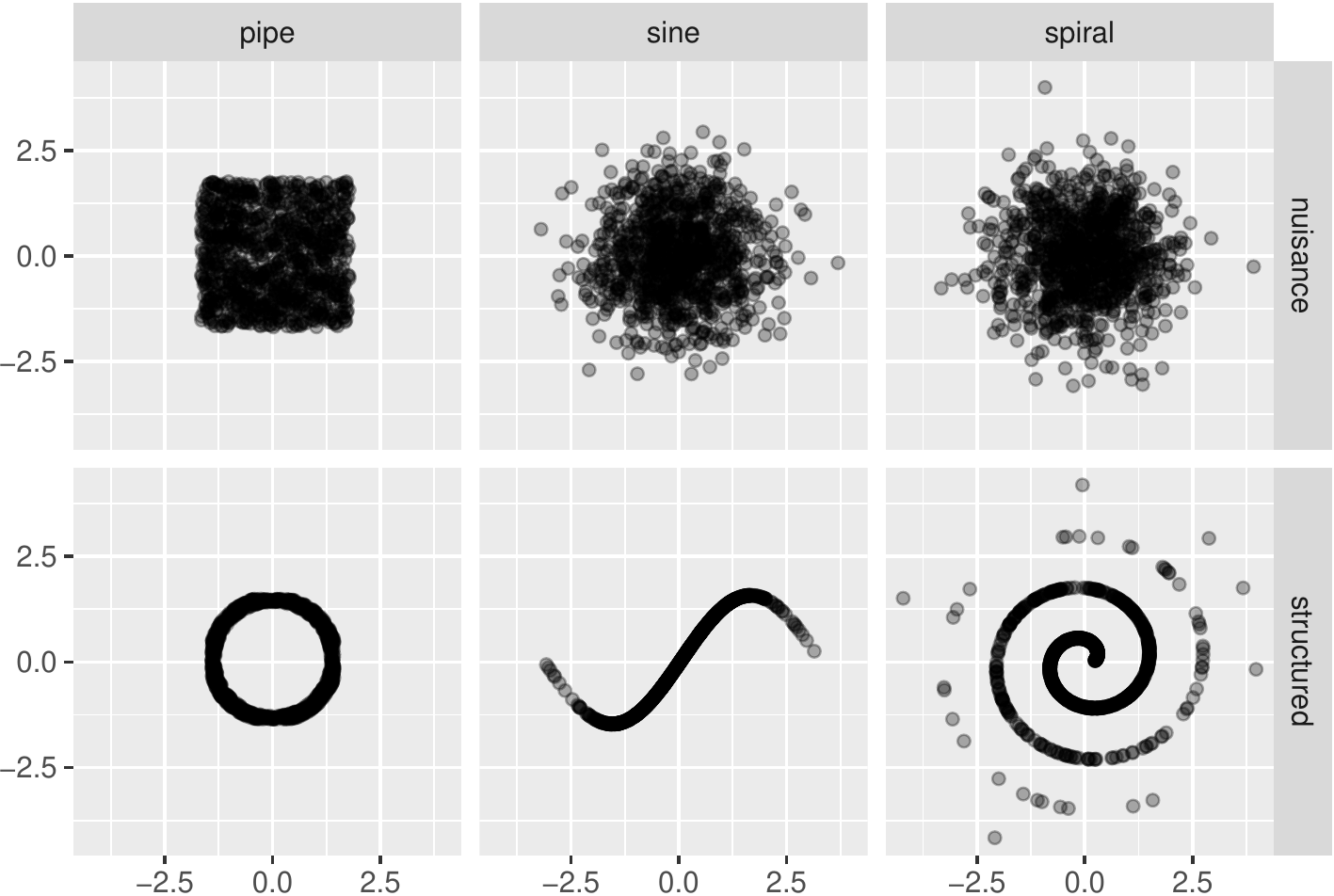} 

}

\caption{Scatterplots of pairs of variables from samples of each family, showing the nuisance variables and structured variables.}\label{fig:data}
\end{figure}

\begin{table}[t]

\caption{\label{tab:indexTable}Comparison of index values between noise projections and structured projections for sample size 1000, using 5th and 95th percentiles from 100 simulated sets. Most indexes have much larger values for the structured projections, except for convex.}
\centering
\begin{tabular}{llrrrrrr}
\toprule
\multicolumn{1}{c}{ } & \multicolumn{1}{c}{ } & \multicolumn{2}{c}{pipe} & \multicolumn{2}{c}{sine} & \multicolumn{2}{c}{spiral} \\
\cmidrule(l{2pt}r{2pt}){3-4} \cmidrule(l{2pt}r{2pt}){5-6} \cmidrule(l{2pt}r{2pt}){7-8}
Index &  & lower & upper & lower & upper & lower & upper\\
\midrule
\addlinespace[-0.2cm]
\multicolumn{8}{l}{\textbf{}}\\
\hspace{1em}holes & noise & 0.37 & 0.46 & 0.00 & 0.01 & 0.00 & 0.01\\
\hspace{1em} & structure & 0.98 & 0.99 & 0.00 & 0.00 & 0.00 & 0.00\\
\addlinespace[-0.2cm]
\multicolumn{8}{l}{\textbf{}}\\
\hspace{1em}convex & noise & 0.70 & 0.73 & 0.57 & 0.69 & 0.54 & 0.68\\
\hspace{1em} & structure & 0.06 & 0.07 & 0.00 & 0.00 & 0.00 & 0.00\\
\addlinespace[-0.2cm]
\multicolumn{8}{l}{\textbf{}}\\
\hspace{1em}skinny & noise & 0.12 & 0.18 & 0.13 & 0.28 & 0.13 & 0.31\\
\hspace{1em} & structure & 0.84 & 0.86 & 0.76 & 0.87 & 0.84 & 0.90\\
\addlinespace[-0.2cm]
\multicolumn{8}{l}{\textbf{}}\\
\hspace{1em}stringy & noise & 0.31 & 0.48 & 0.20 & 0.42 & 0.22 & 0.43\\
\hspace{1em} & structure & 0.57 & 0.74 & 1.00 & 1.00 & 0.88 & 0.98\\
\addlinespace[-0.2cm]
\multicolumn{8}{l}{\textbf{}}\\
\hspace{1em}dcor2D & noise & 0.03 & 0.07 & 0.04 & 0.07 & 0.04 & 0.07\\
\hspace{1em} & structure & 0.17 & 0.18 & 0.96 & 0.98 & 0.14 & 0.17\\
\addlinespace[-0.2cm]
\multicolumn{8}{l}{\textbf{}}\\
\hspace{1em}splines2D & noise & 0.00 & 0.02 & 0.00 & 0.02 & 0.00 & 0.02\\
\hspace{1em} & structure & 0.00 & 0.02 & 1.00 & 1.00 & 0.01 & 0.05\\
\addlinespace[-0.2cm]
\multicolumn{8}{l}{\textbf{}}\\
\hspace{1em}MIC & noise & 0.03 & 0.04 & 0.03 & 0.04 & 0.03 & 0.05\\
\hspace{1em} & structure & 0.56 & 0.58 & 0.98 & 1.00 & 0.40 & 0.45\\
\addlinespace[-0.2cm]
\multicolumn{8}{l}{\textbf{}}\\
\hspace{1em}TIC & noise & 0.02 & 0.02 & 0.02 & 0.02 & 0.02 & 0.02\\
\hspace{1em} & structure & 0.42 & 0.43 & 0.95 & 0.98 & 0.26 & 0.30\\
\bottomrule
\end{tabular}
\end{table}

For simplicity, in the investigations of the index behavior, we fix
\(p=6\), which corresponds to two independent 2D planes containing
nuisance distributions, and one 2D plane containing the structured
distribution. The structured projection is in variables 5 and 6
(\(x_5, x_6\)). Two samples sizes are used: \(n=(100, 1000)\). All
variables are standardized to have mean 0 and standard deviation 1.
Figure \ref{fig:data} shows samples from each of the families, of the
nuisance and structured pairs of variables. Table \ref{tab:indexTable}
compares the PPIs for structured projections against those for nuisance
variables, based on 100 simulated data sets of each type, using sample
size 1000. The lower and upper show the 5th and 95th percentile of
values. The holes index is sensitive only to the pipe distribution. All
other indexes, except convex show distinctly higher values for the
structured projections. The convex index shows the inverse scale to
other indices, thus (1-convex) will be used in the assessment of
performance of PPIs. The scale for the holes index in its original
implementation is smaller than the others ranging from about 0.7 through
1, so it is re-scaled in the performance assessment so that all indices
can be plotted on a common scale of 0-1 (details are given in the
Appendix). Similarly, the TIC index is re-scaled depending on sample
size.

\hypertarget{sec:propAss}{%
\subsubsection{Property assessment}\label{sec:propAss}}

The procedures for assessing the PPI properties of smoothness,
squintability, flexibility, rotation invariance, and speed examined for
samples from the family of data sets are:

\begin{enumerate}
\def\labelenumi{\arabic{enumi}.}
\tightlist
\item
  Compute the PPI values on the tour path along an interpolation between
  pairs of nuisance variables, \(x_1 - x_2\) to \(x_3 - x_4\). The
  result is ideally a smooth change in low values. This checks the
  smoothness property.
\item
  Change to a tour path between a pair of nuisance variables
  \(x_1 - x_2\) and the structured pair of variables \(x_5 - x_6\) via
  the intermediate projection onto \(x_1 - x_5\), and compute the PPI
  along this. This examines the squintability, and smoothness. If the
  function is smooth and slowly increases towards the structured
  projection, then the structure is visible from a distance.
\item
  Use the guided tour to examine the ease of optimization. This depends
  on having a relatively smooth function, with structure visible from a
  distance. One index is optimized to show how effectively the maximum
  is attained, and the values for other PPIs is examined along the same
  path, to examine the similarity between PPIs.
\item
  Rotation invariance is checked by computing PPIs on rotations of the
  structured projection.
\item
  Computational speed for the selected indexes is examined on a range of
  sample sizes.
\end{enumerate}

\hypertarget{ppi-traces-over-a-tour-sequence-of-interpolated-nuisance-projections}{%
\subsection{PPI traces over a tour sequence of interpolated nuisance
projections}\label{ppi-traces-over-a-tour-sequence-of-interpolated-nuisance-projections}}

\label{sec:smooth}

\begin{figure}

{\centering \includegraphics[width=\textwidth]{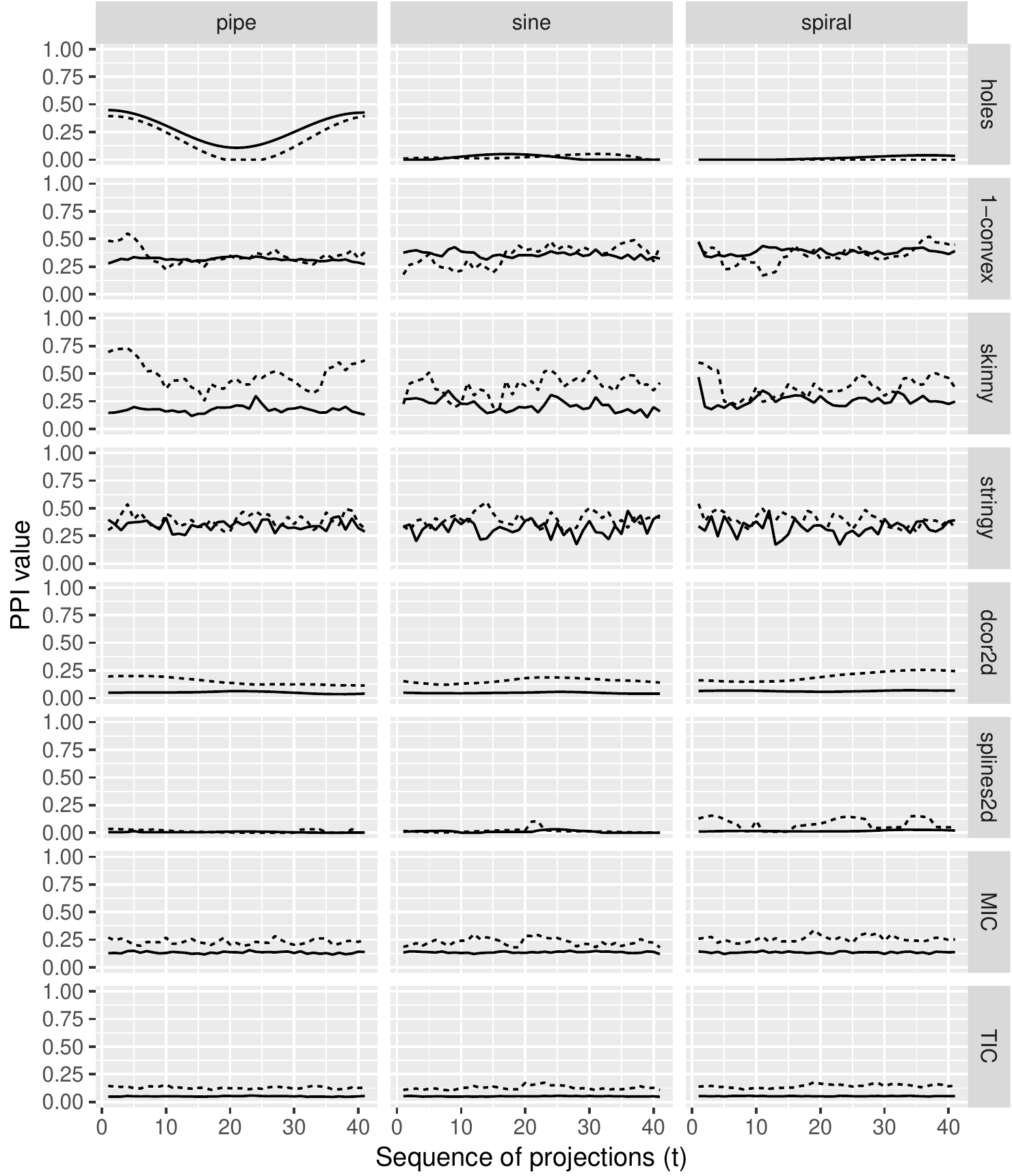} 

}

\caption{PPIs for  projections along a geodesic interpolation between two nuisance projections. All projections would be nuisance so the PPI are ideally low and smooth, with little difference between sample sizes (solid lines: $n=1000$; dashed: $n=100$). The scagnostic PPIs are noisy. Some indexes have distinct differences in values between sample sizes. (This is not an optimization path, but an interpolation containing 41 projections between two known projections.)}\label{fig:plotV1V2toV3V4}
\end{figure}

Figure \ref{fig:plotV1V2toV3V4} shows the traces representing the index
values calculated across a tour between a pair of nuisance projections.
The tour path is generated by interpolating between the two independent
nuisance planes, i.e.~from the projection onto \(x_1\)-\(x_2\) to one
onto \(x_3\)-\(x_4\). The range of each axis is set to be the limits of
the index, as might be expected over many different data sets, 0 to 1.
Each projection in the interpolation will also be noise. Two different
sample sizes are show, \(n=100\) as a dashed line, and \(n=1000\) as a
solid line. The ideal trace is a smooth function, with relatively low
values, and no difference between the sample sizes. A major feature to
notice is that the scagnostics produce noisy functions, which is
problematic, because small changes in the projection result in big jumps
in the value. This will make them difficult to optimize. On the other
hand holes, dcor2d, splines2d, MIC and TIC are relatively smooth
functions.

Several of the indexes are sensitive to sample size also, the same
structured projection with differing numbers of points, produces
different values.

\hypertarget{ppi-traces-over-a-tour-sequence-between-nuisance-and-structured-projections}{%
\subsection{PPI traces over a tour sequence between nuisance and
structured
projections}\label{ppi-traces-over-a-tour-sequence-between-nuisance-and-structured-projections}}

\label{sec:squintability}

\begin{figure}

{\centering \includegraphics[width=\textwidth]{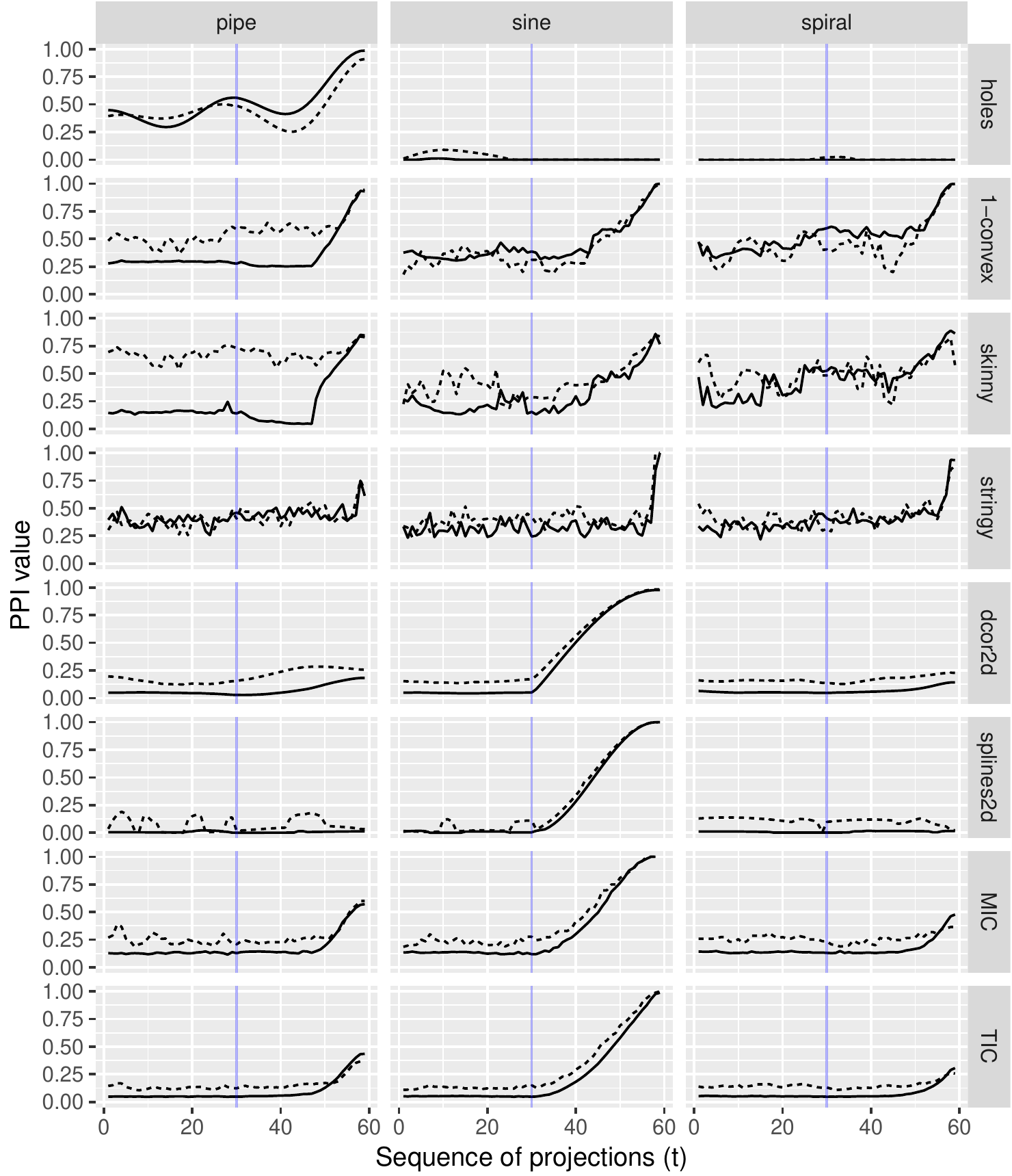} 

}

\caption{PPIs for projections along an interpolation between nuisance and structured projections, following $x_1$-$x_2$ to $x_1$-$x_5$ to $x_5$-$x_6$ (solid lines: $n=1000$; dashed: $n=100$). The vertical blue line indicates the position of the projection onto $x_1$-$x_5$ in the sequence. Peaks at the end of the sequence indicate the index sees the structure. The scagnostics, MIC and TIC see all three structures, so are more flexible for general pattern detection. Holes only responds to the pipe, and is a multimodal function for this data with a local maximum at $x_1$-$x_5$. (This is not an optimization path, but an interpolation containing 59 projections between three known projections.) }\label{fig:wIntermediate}
\end{figure}

Figure \ref{fig:wIntermediate} shows the PPIs for a tour sequence
between a nuisance and structured projection. A long sequence is
generated where the path interpolates between projections onto
\(x_1\)-\(x_2\), \(x_1\)-\(x_5\), \(x_5\)-\(x_6\), in order to see some
of the intricacies of holes index. Sample size is indicated by line
type: dashed being \(n=100\) and solid is \(n=1000\). The beginning of
the sequence is the nuisance projection and the end is the structured
projection. The index values for most PPIs increases substantially
nearing the structured projection, indicating that they ``see'' the
structure. Some indexes see all three structures: scagnostics, MIC and
TIC, which means that they are flexible indexes capable of detecting a
range of structure. (Grimm 2016)'s indexes, dcor2d and splines2d, are
excellent for detecting the sine, and they can see it from far away,
indicated by the long slow increase in index value. The holes index
easily detects the pipe, and can see it from a distance, but also has
local maxima along the tour path. The scagnostic index, stringy, can see
the structure but is myopic, only when it is very close. Interestingly
the scagnostic, skinny, sees the spiral from a distance.

\hypertarget{optimization-check-with-the-guided-tour}{%
\subsection{Optimization check with the guided
tour}\label{optimization-check-with-the-guided-tour}}

\label{sec:guided}

Before applying the new index functions, with the guided tour on real
examples, we test them on the simulated dataset to understand the
performance of the optimization. The guided tour combines optimization
with interpolation between pairs of planes. Target planes of the path
are chosen to maximize the PPI. There are three derivative-free
optimization methods available in the guided tour:
\verb# search_better_random# (1), \verb# search_better# (2), and
\verb# search_geodesic# (3). Method 1 casts a wide net randomly
generating projection planes, computing the PPIs and keeping the best
projection, and method 2 conducts a localized maximum search. Method 3
is quite different: a local search is conducted to determine a promising
direction, and then this direction is followed until the maximum in that
direction is found. For all methods the optimization is iterative, the
best projections form target planes in the tour, the tour path is the
interpolation to this target, and then a new search for a better
projection is made, followed by the interpolation. For each projection
during the interpolation steps, the PPI is recorded.

The stopping rule is that no better projections are found after a fixed
number of tries, given a fixed tolerance value measuring difference. For
method 1 and 2 two additional parameters control the optimization: the
search window \(\alpha\), giving the maximum distance from the current
plane in which projections are sampled, and the cooling factor, giving
the incremental decrease in search window size. Method 3 in principle
also has two free parameters, which are however fixed in the current
implementation. The first is the small step size used when evaluating
the most promising direction, it is fixed to 0.01, and the second
parameter being the window over which the line search is performed,
fixed to \(\pm \pi/4\) away from the current plane.

For distributions and indexes with smooth behavior and good
squintability, method 3 is the most effective method for optimization.
If these two criteria are not met the method may still be useful, but
only given an informed starting projection. In such cases we can follow
a method similar to that proposed by Friedman (1987): we break the
optimization in two distinct steps. A first step (``scouting'') uses
method 2 with large search window and no cooling as a way of stepping
over fluctuations and local maxima and yielding an approximation of the
global maximum. Note that this likely requires large number of tries,
especially as dimension increases, since most randomly picked planes
will not be interesting. The second step uses method 3 starting from the
approximate maximum, which will take small steps to refine the result to
be closer to the global maximum.

\hypertarget{looking-down-the-pipe}{%
\subsubsection{Looking down the pipe}\label{looking-down-the-pipe}}

Despite the simple structure, the pipe is relatively difficult for the
PPIs to find. For the TIC index, there is a fairly small squint angle.
For the holes index, there are several local maxima, that divert the
optimizer. There is a hint of this from Figure \ref{fig:wIntermediate}
because the initial projection (left side of trace) of purely noise
variables has a higher index value than the linear combinations of noise
and structured variables along the path. The uniform distribution was
used to generate the noise variables, which has a higher PPI value than
a normal distribution, yielding the higher initial value. In addition, a
local maximum is observed whenever the pair of variables is one
structured variable and one noise variable, because there is a lighter
density in the center of the projection.

The optimization is done in two stages, a scouting phase using method 2,
and a refinement stage using method 3. For the scouting we use
\(\alpha = 0.5\) and stopping condition of maximum 5000 tries, and we
optimized the TIC index.

\begin{figure}

{\centering \includegraphics[width=0.8\textwidth]{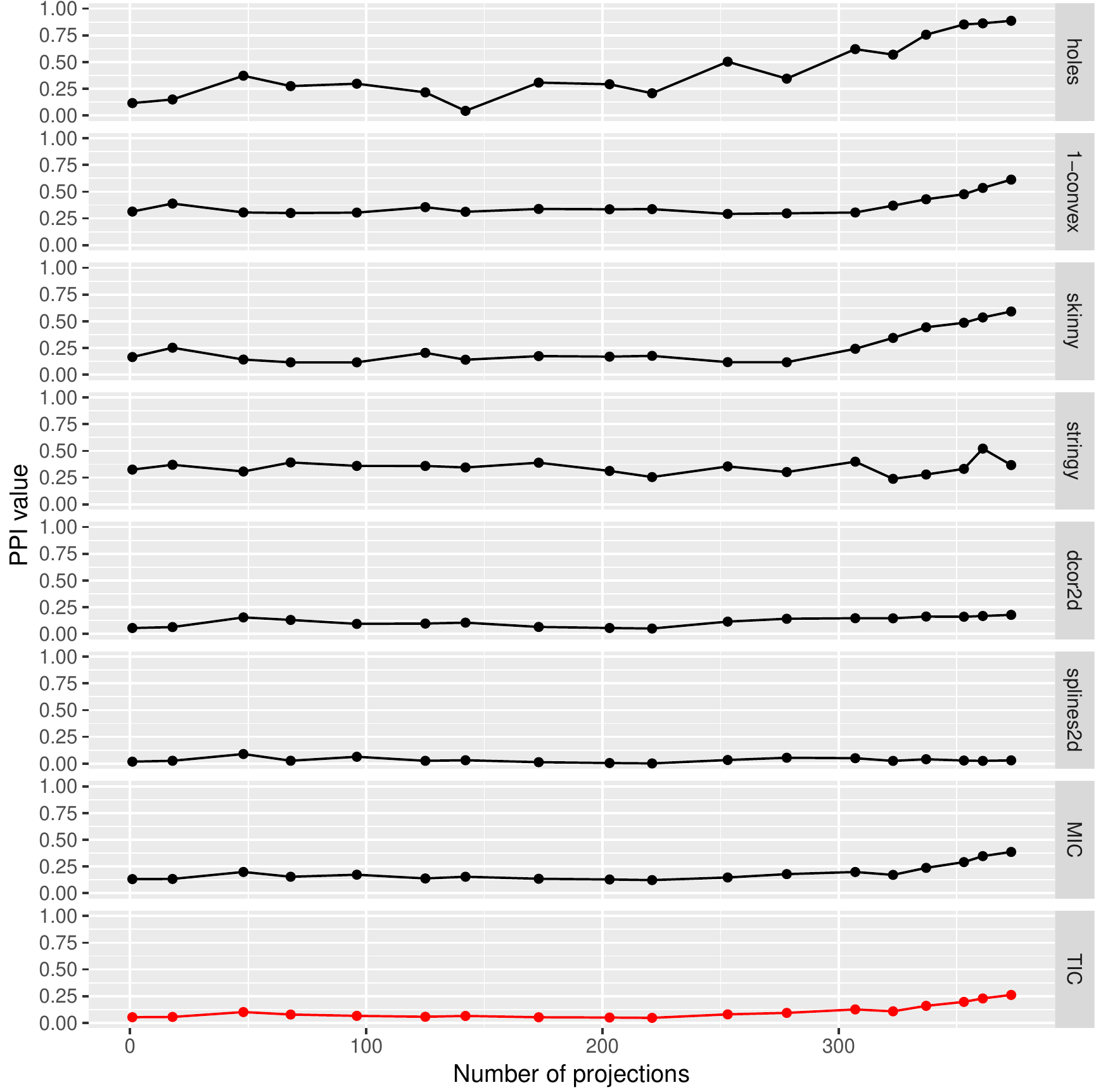} 

}

\caption{PPIs for a sequence of projections produced by scouting for the pipe using optimization method 2 on the TIC index. Other PPI values are shown for interest. Only the values on the target planes are shown. Despite the small maximum value of TIC for this data, it identifies the pipe.}\label{fig:pipeFirstRun}
\end{figure}

Figure \ref{fig:pipeFirstRun} shows the target projections (points)
selected during the scouting with method 2 on the TIC index. The focus
is on the target projections rather than the interpolation between them,
because the optimization is done off-line, and only the targets are used
for the next step. The horizontal distance between the points in the
plot reflects the relative geodesic distance between the planes in the
6D space. All of the other indexes are shown for interest. The TIC index
value is generally low for this data, although it successfully detects
the pipe. The holes, convex, skinny, and to some extent MIC, mirror the
TIC performance. The holes differs in that it has some intermediate high
values which are likely the indication of multi-modality of this index
on this data.

The final views obtained in each of the two stages are compared in the
Appendix.

\hypertarget{finding-sine-waves}{%
\subsubsection{Finding sine waves}\label{finding-sine-waves}}

\begin{figure}

{\centering \includegraphics[width=0.8\textwidth]{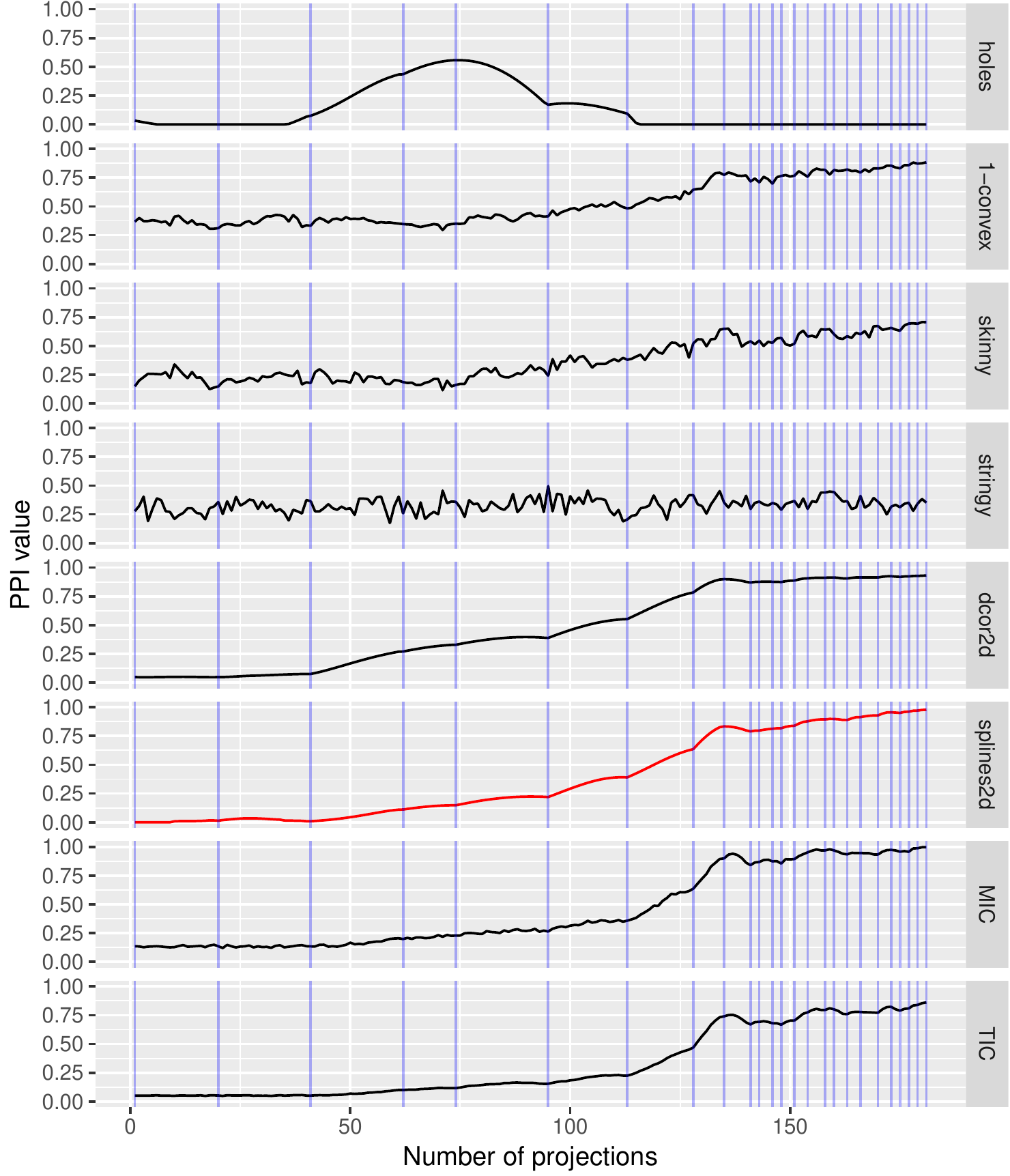} 

}

\caption{PPIs for sequence of projections produced by a guided tour optimizing  the splines2d index, using optimization method 3, for the sine data, with $n=1000$. Anchor planes are marked by the blue vertical lines, and are closer to each other approaching the maxima. The sine is found relatively easily, by splines2d, and it is indicated that MIC, TIC, dcor2d and convex would also likely find this structure.}\label{fig:findsine}
\end{figure}

Given the patterns in Figure \ref{fig:wIntermediate} it would be
expected that the sine could be found easily, using only optimization
method 3 with the splines2d, dcor2d, MIC or TIC indexes. This is
examined in Figure \ref{fig:findsine}. Optimization is conducted using
the splines2d index, and the trace of the PPI over the optimization is
shown, along with the PPI values for the other indexes over that path.
The vertical blue lines indicate anchor bases, where the optimizer
stops, and does a new search. The distance between anchor planes is
smaller as the maximum is neared.

The only complications arise from a lack of rotation invariance of the
splines2d index. It is not easily visible here, but it is possible that
the best projection will have a higher PPI. The index changes depending
on the basis used to define the plane, but the geodesic interpolation
conducted by the tour uses any suitable basis to describe the plane,
ignoring that which optimizes the PPI. This is discussed in section
\ref{sec:rot}.

\hypertarget{spiral-detection}{%
\subsubsection{Spiral detection}\label{spiral-detection}}

The spiral is the most challenging structure to detect because it has a
small squint angle (Posse 1995b), especially as the ratio of noise to
structure dimensions increases. This is explored using optimization
method 2 to scout the space for approximate maxima. The skinny
scagnostic index is used because it was observed (Figure
\ref{fig:wIntermediate}) to be sensitive to this structure, although the
noisiness of the index might be problematic. The stringy appears to be
more sensitive to the spiral, but it has a much smaller squint angle.

\begin{figure}

{\centering \includegraphics[width=0.9\textwidth]{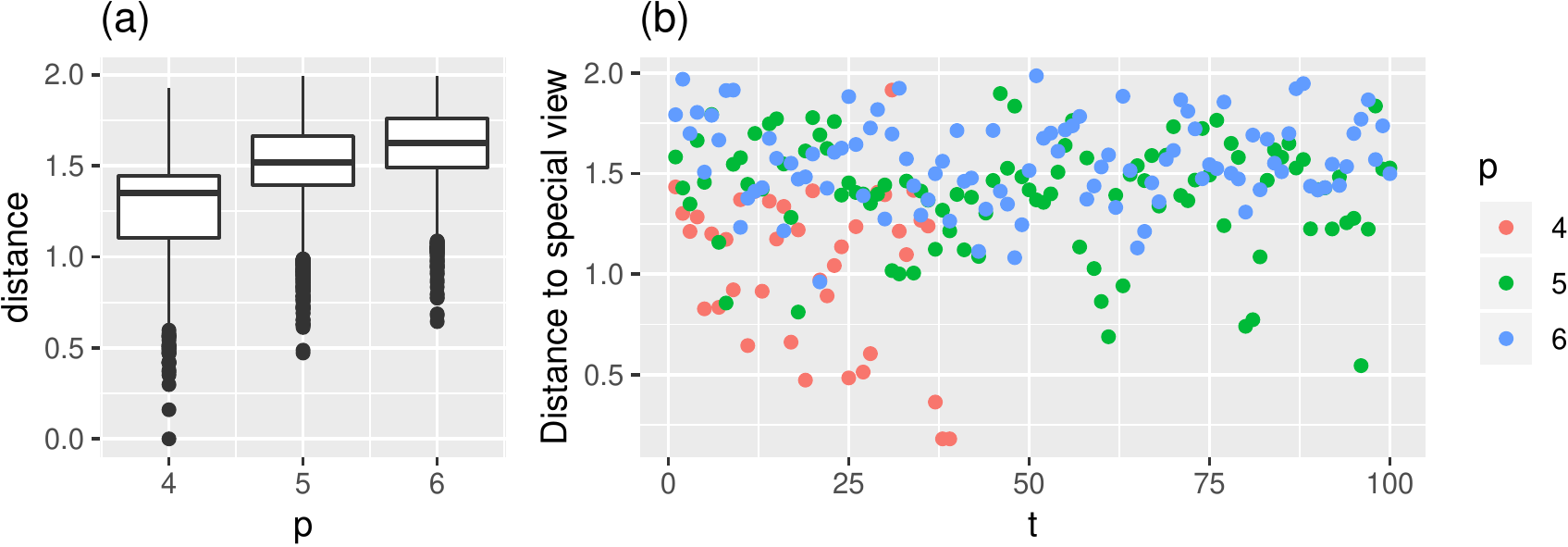} 

}

\caption{Guided tour optimizing the skinny index for the Sprial dataset with 1000 datapoints, with p = 4, 5, 6. The left plot shows the distribution of pairwise distances between planes obtained via the guided tour, the right shows the evolution of distance to the ideal plane as the index is being optimized.}\label{fig:findspiral}
\end{figure}

The search is conducted for \(p=4,5,6\) which would correspond to 2, 3
and 4 noise dimensions respectively. In addition we examine the distance
between planes, using a Frobenius norm, as defined by Equation 2 of Buja
et al. (2005), and available in the \texttt{proj\_dist} function in the
tourr package, to compare searches across dimensions. The distance
between planes is related to squint angle, how far away from the ideal
projection can the structure be glimpsed. We estimate the squint angle
depending on the number of noise dimensions in the appendix. In order
for the optimizer to find the spiral, the distance between planes would
need to be smaller than the squint angle. Figure \ref{fig:findspiral}
summarizes the results. When \(p=4\) the scouting method effectively
finds the spiral. Plot (a) shows the side-by-side boxplots of pairwise
distances between planes examined during the optimization, for
\(p=4,5,6\). These are on average smaller for the lower dimension, and
gradually increase as dimension increases. This is an indication of the
extra computation needed to brute force find the spiral as noise
dimensions increase. Plot (b) shows the distance of the plane in each
iteration of the optimization to the ideal plane, where it can be seen
that only when \(p=4\) does it converge to the ideal. Its likely that
expanding the search space should result in uncovering the spiral in
higher dimensions, which however requires tuning of the stopping
conditions and long run times.

\hypertarget{rotational-invariance-or-not}{%
\subsection{Rotational invariance or
not}\label{rotational-invariance-or-not}}

\label{sec:rot}

\begin{figure}

{\centering \includegraphics[width=.6\textwidth]{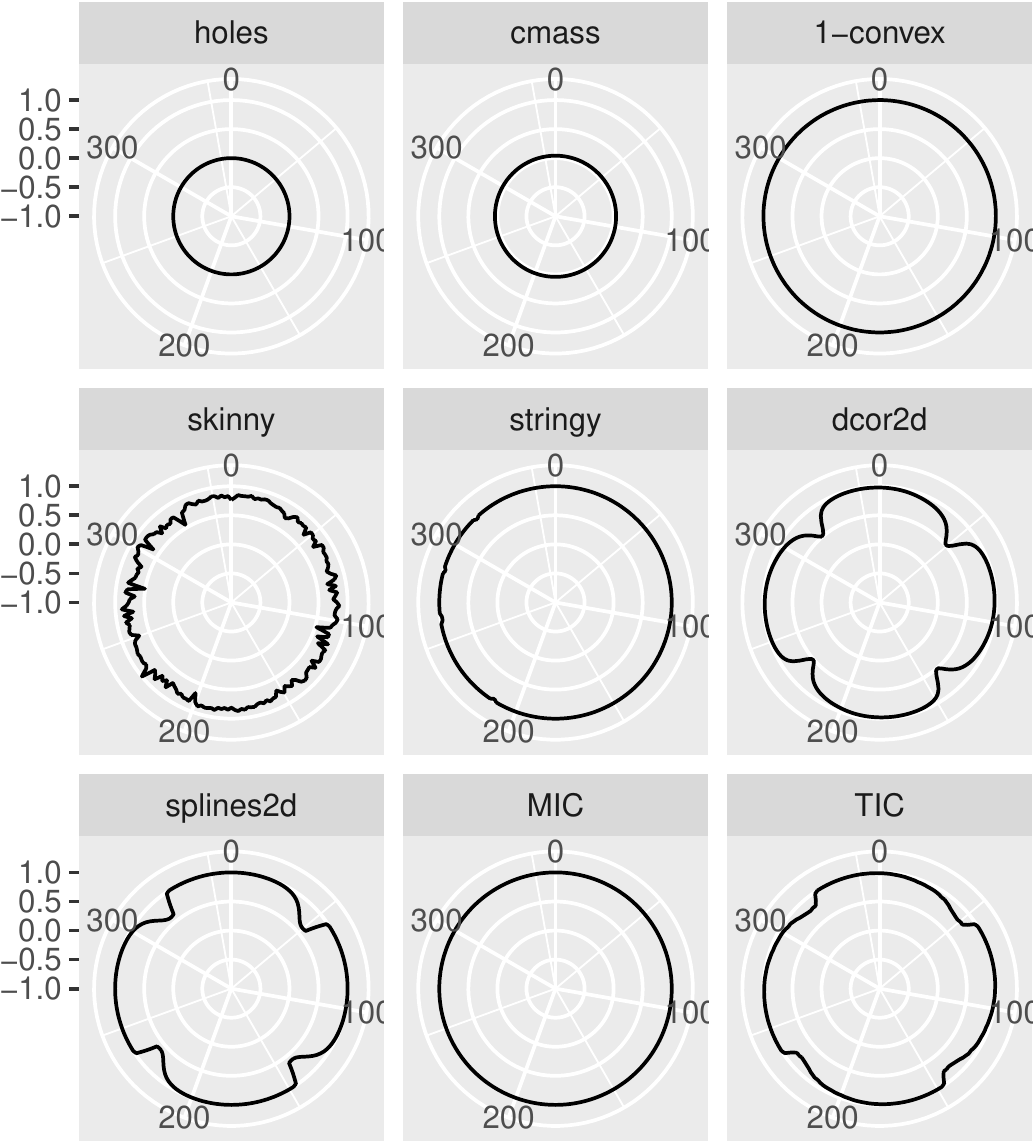} 

}

\caption{PPI for rotations of the sine 1000 data, to examine rotation invariance. Most are close to rotation invariant, except for skinny, dcor2d, splines2d and TIC.}\label{fig:rotationDep}
\end{figure}

Rotational invariance is examined using the sine data (\(x_5\)-\(x_6\)),
computing PPI for different rotations within the 2D plane, parameterized
by angle. Results are shown in Figure \ref{fig:rotationDep}. Several
indexes are invariant, holes, convex and MIC, because their value is
constant around rotations. The dcor2d, splines2d and TIC index are
clearly not rotationally invariant because the value changes depending
on the rotation. The scagnostics indexes are approximately rotationally
invariant, but particularly the skinny index has some random variation
depending on rotation.

\hypertarget{speed}{%
\subsection{Speed}\label{speed}}

Examining the computing time as a function of sample size we find that
scagnostics and splines2d are fast even for large samples, while all
other index functions slow rapidly with increasing sample size. Detailed
results are shown in the Appendix.

\hypertarget{parameter-choices}{%
\subsection{Parameter choices}\label{parameter-choices}}

Some PPIs have a choice of parameters, and the choice can have an effect
on function smoothness, and sensitivity to structure. In the Appendix we
examine the dependence of the scagnostics indexes on the binning,
showing that even with small number of bins the indexes are noisy, while
they lose ability to see structure. Sensitivity to the number of bins in
the MIC index is also examined, showing that tuning the parameter can
improve the final result.

\hypertarget{index-enhancement}{%
\subsection{Index enhancement}\label{index-enhancement}}

We identified two potential improvements. First, the issue of noisy
index functions may be addressed via smoothing, and we explore different
smoothing options for the examples of the skinny and stringy index in
the Appendix. In addition, rotation dependent indexes may be enhanced by
redefining them in a rotation invariant way.

\hypertarget{summary}{%
\subsection{Summary}\label{summary}}

Our results can be summarized by evaluating and comparing the advantages
and disadvantages of each index function according to the criteria
presented above. Such an overview is given in Table \ref{tab:summary},
listing if the criteria is fully met (\checkmark), there are some
shortcomings (\(\cdot\)) or failure (\(\times\)). (The holes index does
not appear in the summary because its performance understood, and is not
being examined here.) We find that none of the indexes considered meet
all criteria, and in particular rotation invariance is often not
fulfilled. In addition the limited flexibility of most indexes
highlights the importance of index selection in the projection pursuit
setup. Table \ref{tab:summary} further suggests that there is much room
for the improvement of index functions detecting unusual association
between model parameters.

\begin{table}
\begin{center}
\caption{Summary of findings, showing to what extend the considered new index functions pass the criteria for a good PPI. ``\checkmark" symbols good behavior, ``$\cdot$" symbols some issues while ``$\times$" symbols failure on the corresponding criteria. Each index has particular strengths and drawbacks and selection must be guided by the considered example, see text for details.}
\label{tab:summary}
\begin{tabular}{|l|c|c|c|c|c|c|}
\hline
Index & smooth & squintability & flexible & rotation invariant & speed \\
\hline
convex & $\cdot$ & $\cdot$ & $\cdot$ & \checkmark & \checkmark \\
skinny & $\cdot$ & $\cdot$ & $\cdot$ & $\times$ & \checkmark \\
stringy & $\times$ & $\times$ & $\cdot$ & $\cdot$ & \checkmark \\
\hline
splines2D & \checkmark & \checkmark & $\cdot$ & $\times$ & \checkmark \\
dcor2D & \checkmark & \checkmark & $\cdot$ & $\times$ & $\cdot$ \\
\hline
MIC & $\cdot$ & \checkmark  & \checkmark & $\cdot$ & $\cdot$ \\
TIC & $\cdot$ & \checkmark  & \checkmark & $\cdot$ & $\cdot$ \\
\hline
\end{tabular}
\end{center}
\end{table}

\hypertarget{application-to-physics-examples}{%
\section{Application to physics
examples}\label{application-to-physics-examples}}

\label{sec:phys}

This section describes the application of these projection pursuit
indices to find two-dimensional structure in two multidimensional
gravitational waves problems.

The first example contains 2538 posterior samples obtained by fitting
source parameters to the observed gravitational wave signal GW170817
from a neutron star merger (Abbott and others 2017). Data has been
downloaded from (``LIGO'' 2018). The fitting procedures are described in
detail in Abbott and others (2018). We consider six parameters of
physical interest (6-D) with some known relationships. Projection
pursuit is used to find the known relationships.

The second example contains data generated from a simulation study of a
binary black hole (BBH) merger event, as described in Smith et al.
(2016). There are 12 parameters (12-D), with multiple nuisance
parameters. Projection pursuit uncovers new relationships between
parameters.

\hypertarget{neutron-star-merger}{%
\subsection{Neutron star merger}\label{neutron-star-merger}}

A scatterplot matrix (with transparency) of the six parameters is shown
in the Appendix. (In astrophysics, scatterplot matrices are often called
``corner plots'' (Foreman-Mackey 2016).) The diagonal shows a univariate
density plot of each parameter, and the upper triangle of cells displays
the correlation between variables. From this it can be seen that m1 and
m2 are strongly, and slightly, nonlinearly associated. Between the other
variables we observe some linear association (R1, R2), some nonlinear
association (L1, L2, R1, R2), heteroskedastic variance in most variables
and some bimodality (R1, L1, L2, m1, m2).

The model describes a neutron star merger and contains 6 free
parameters, with each neutron star described by its mass \(m\) (m1, m2)
and radius \(R\) (R1, R2), and a so-called tidal deformability parameter
\(\Lambda\) (L1, L2) which is a function of the mass and radius,
approximately proportional to \((m/R)^{-5}\).

\hypertarget{data-pre-pocessing}{%
\subsubsection{Data pre-pocessing}\label{data-pre-pocessing}}

Because m1 and m2 are very strongly associated, m2 is dropped before
doing PP. This relationship is obvious from the scatterplots of pairs of
variables and does not need to be re-discovered by PP.

All variables are scaled to range between 0 and 1. The purpose is that
range differences in individual variables should not affect the
detection of relationships between multiple variables. Standardizing the
range will still leave differences between the standard deviations of
the variables, and for this problem this is preferred. Differences in
the standard deviations can be important for keeping the non-linear
relationships visible to PP.

\hypertarget{applying-pp}{%
\subsubsection{Applying PP}\label{applying-pp}}

With only five parameters, a reasonable start is to examine the 5D space
using a grand tour. This quickly shows the strong nonlinear
relationships between the parameters. PP is then used to extract these
relationships. The best index for this sort of problem is the splines2d,
and it is fast to compute.

\begin{figure}

{\centering \includegraphics[width=\textwidth]{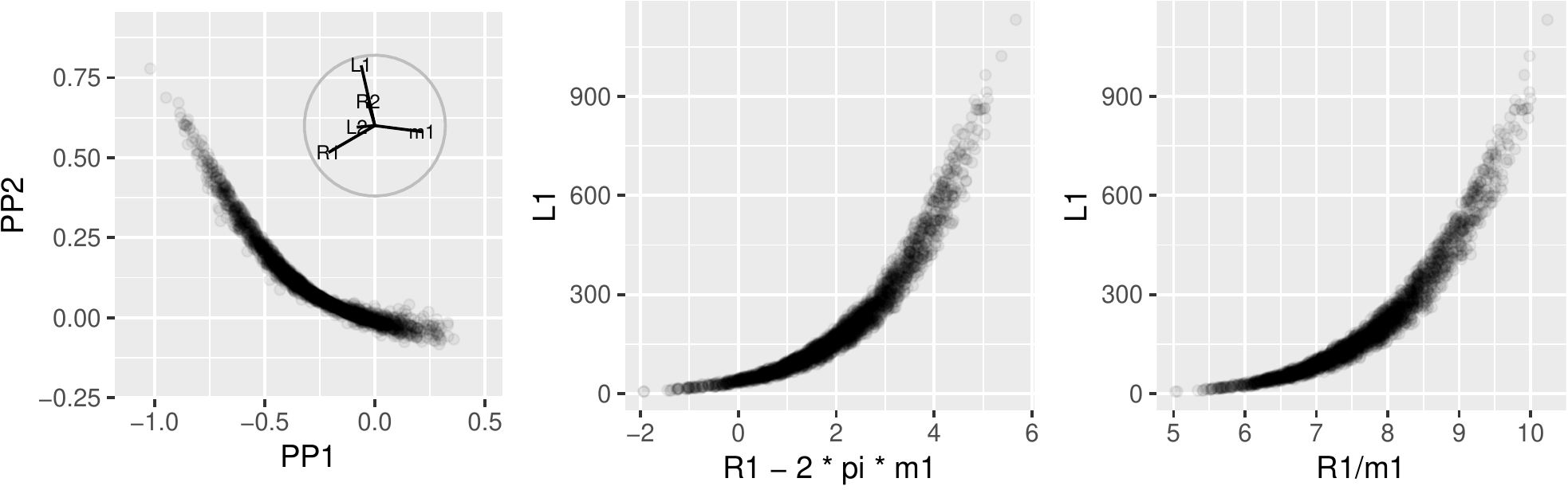} 

}

\caption{Comparison of guided tour final view (left), approximation based on original parameters (middle) and expected relation based on analysis setup (right).}\label{fig:nsePlotOrig}
\end{figure}

Figure \ref{fig:nsePlotOrig} shows the optimal projection found by
splines2d, a reconstructed view obtained by manually combining
parameters, and a plot of the known relationship between parameters.

\begin{figure}

{\centering \includegraphics[width=\textwidth]{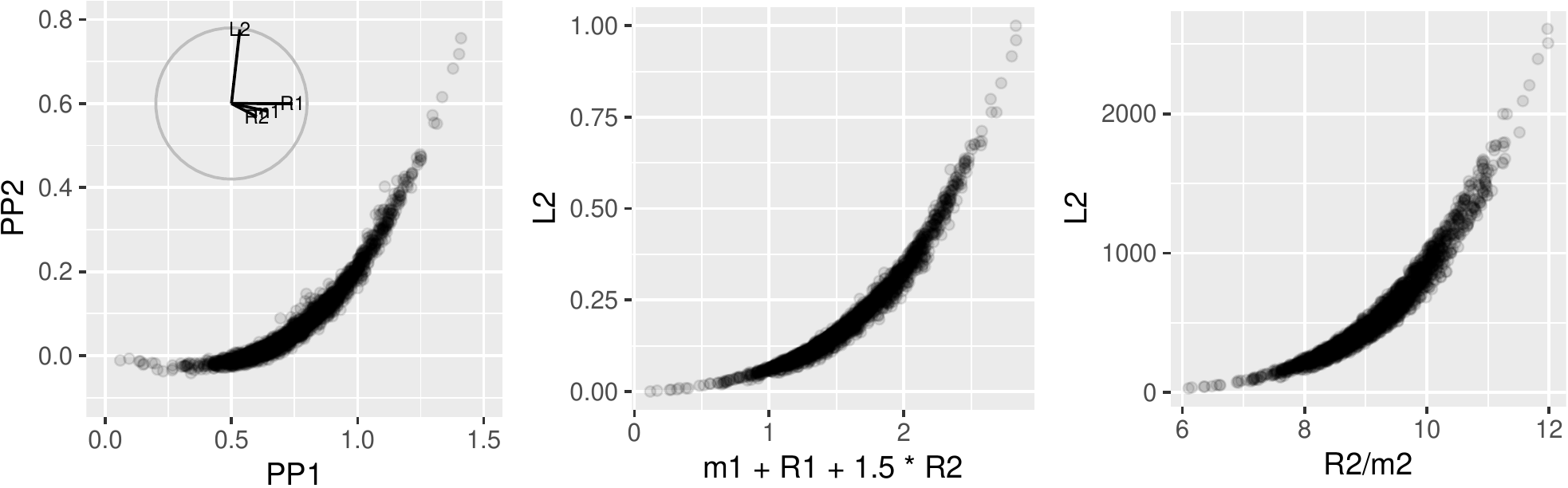} 

}

\caption{Removing L1 and optimize again of the remaining parameters, where m2 remains removed from the set. Because of parameter correlations we can recover clear description of L2 as a function of the other parameters, despite m2 missing.}\label{fig:nseRemL1}
\end{figure}

To further investigate relationships between parameters, \(L1\) is
removed and PP with the splines2D is applied to the remaining four
parameters. The dependence of \(L2\) on the mass and radius of the
lighter neutron star, is revealed (Figure \ref{fig:nseRemL1} left plot).
A manual reconstruction shows this is a relationship between L2, R1, R2
and m1 (middle plot), but it is effectively the known relationship
between L2, R2 and m2 (right plot) -- m2 is latently in the relationship
though m1.

\hypertarget{black-hole-simulation}{%
\subsection{Black hole simulation}\label{black-hole-simulation}}

This data contains posterior samples from simulation from a model
describing a binary black hole (BBH) merger event. There are twelve
model parameters. Flat priors are used for most model parameters.

A scatterplot matrix, of nine of the twelve parameters, is shown in the
Appendix. (Parameter m2 is not shown because it is strongly linearly
associated with m1, phi\(\_\)jl and psi are not shown because they are
uniform, and not associated with other parameters.) Among the nine
plotted parameters, strong nonlinear relationships can be seen between
the parameters ra, dec and time. The first two describe the position of
the event in the sky, and time is the merging time (in GPS units).
Because of the elliptical relationship between dec and time, the TIC
index is used for PP, even though it is slow to compute. Between the
other parameters, the main structure seen is multimodality and some
skewness. These patterns are representative of the likelihood function,
since most priors are flat, or built to capture growth with volume
rather than distance.

\hypertarget{data-pre-processing}{%
\subsubsection{Data pre-processing}\label{data-pre-processing}}

The analysis is conducted on 11 of the twelve parameters. One variable
is removed, m2, because it is so strongly associated with m1. All
parameters are scaled into the range 0 to 1.

\hypertarget{applying-pp-1}{%
\subsubsection{Applying PP}\label{applying-pp-1}}

\hypertarget{exploring-11d-with-all-pp-indexes}{%
\paragraph{Exploring 11D with all PP
indexes}\label{exploring-11d-with-all-pp-indexes}}

\begin{figure}

{\centering \includegraphics[width=\textwidth]{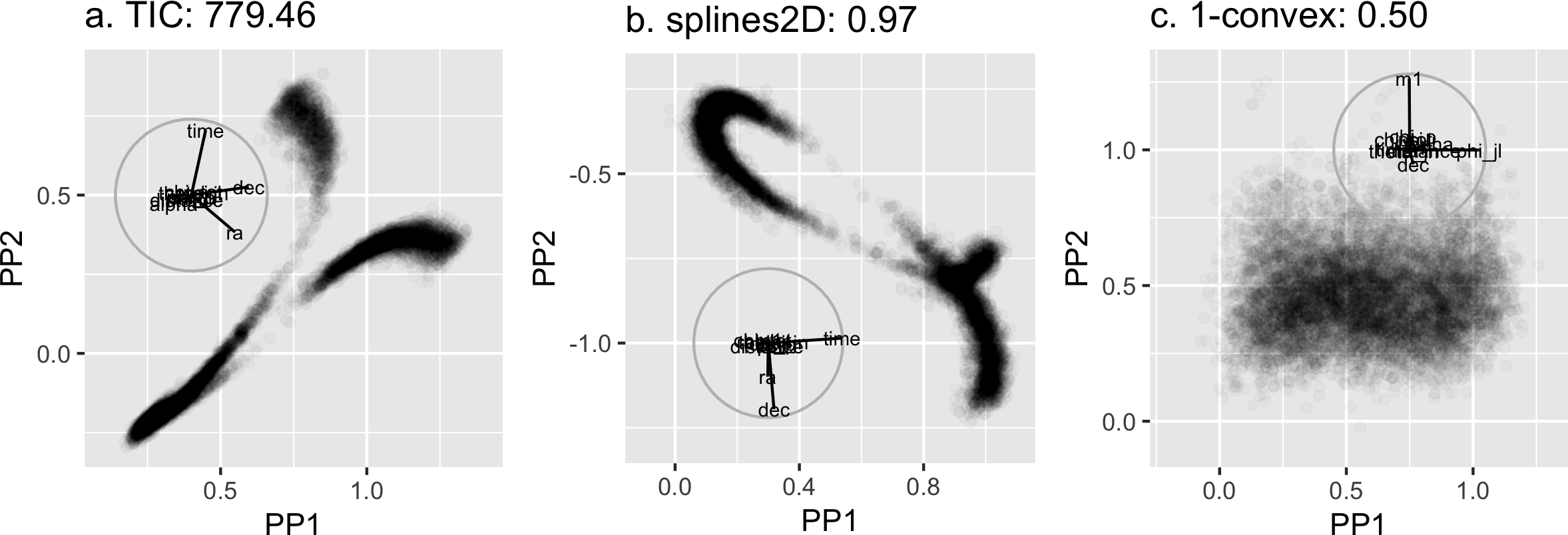} 

}

\caption{Projections corresponding to the maxima of three indices: TIC, splines2D and 1-convex. Projections a, b found by TIC and splines2d are very similar, and involve the same three parameters, ra, dec and time. The 1-convex index finds a very different view.}\label{fig:bbhGuided}
\end{figure}

All seven PP indexes are applied to the data. Figure \ref{fig:bbhGuided}
shows the projections that maximize three of the indexes. TIC and
splines2d indexes identify very similar projections, that are based on
the three parameters, dec, time and ra. This is to be expected based on
the pairwise scatterplots. On the other hand, the 1-convex index finds a
very different view, but this is because the optimization doesn't
adequately reach a maximum for this index.

\hypertarget{exploring-reduced-space}{%
\paragraph{Exploring reduced space}\label{exploring-reduced-space}}

\begin{figure}

{\centering \includegraphics[width=\textwidth]{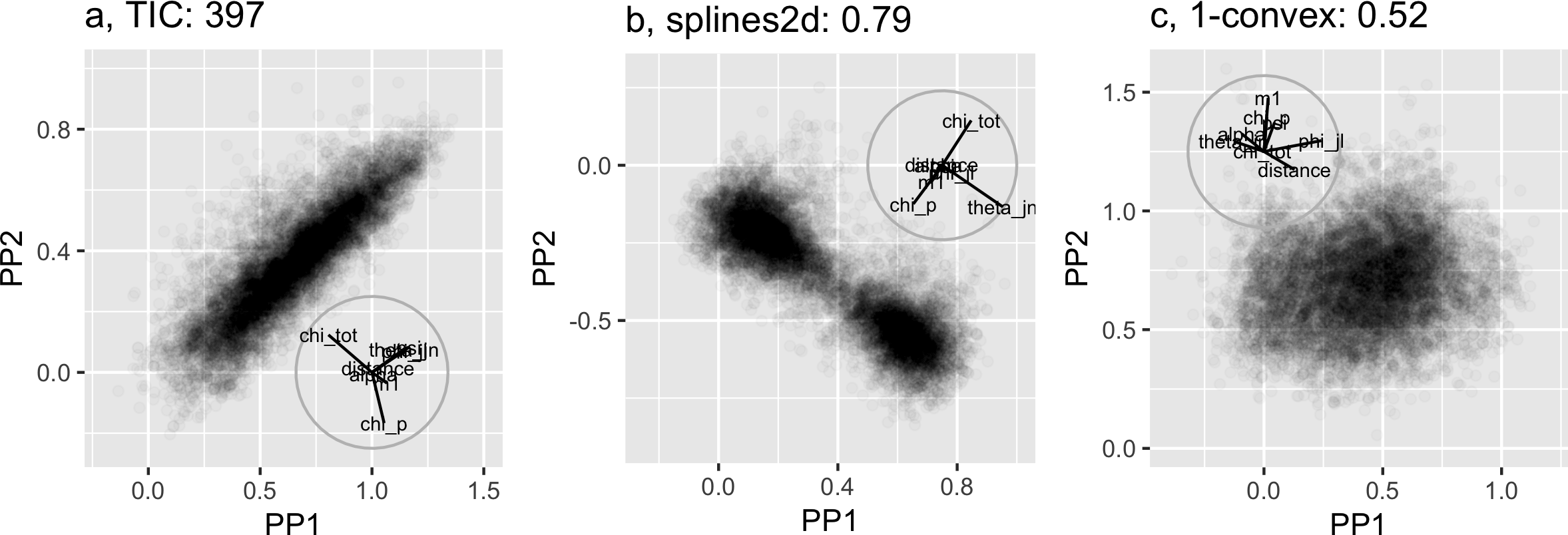} 

}

\caption{Projections of the reduced 8D space corresponding to the maxima of three indices: TIC, splines2d and 1-convex.}\label{fig:bbhGuided2}
\end{figure}

The variables time, dec and ra are dropped from the data, and PP is
applied to the remaining 8D space. Figure \ref{fig:bbhGuided2} shows the
projections which maximize the TIC, splines2D and 1-convex indices. The
results provide similar information as already learned from the
scatterplot matrix. The parameters chi\(\_\)tot and chi\(\_\)p are
linearly related (TIC maxima), and theta\(\_\)jn has a bimodal
distribution yielding the figure 8 shape found by the splines2d index.
The 1-convex index finds nothing interesting.

\hypertarget{effect-of-random-starts-and-subsets-used}{%
\paragraph{Effect of random starts, and subsets
used}\label{effect-of-random-starts-and-subsets-used}}

The initial conditions for the optimization, and the subset of variables
used, can have a large effect on the projections returned. We illustrate
this using only the splines2d index, and find that there is one more
association that can be learned that was masked earlier.

\begin{figure}

{\centering \includegraphics[width=\textwidth]{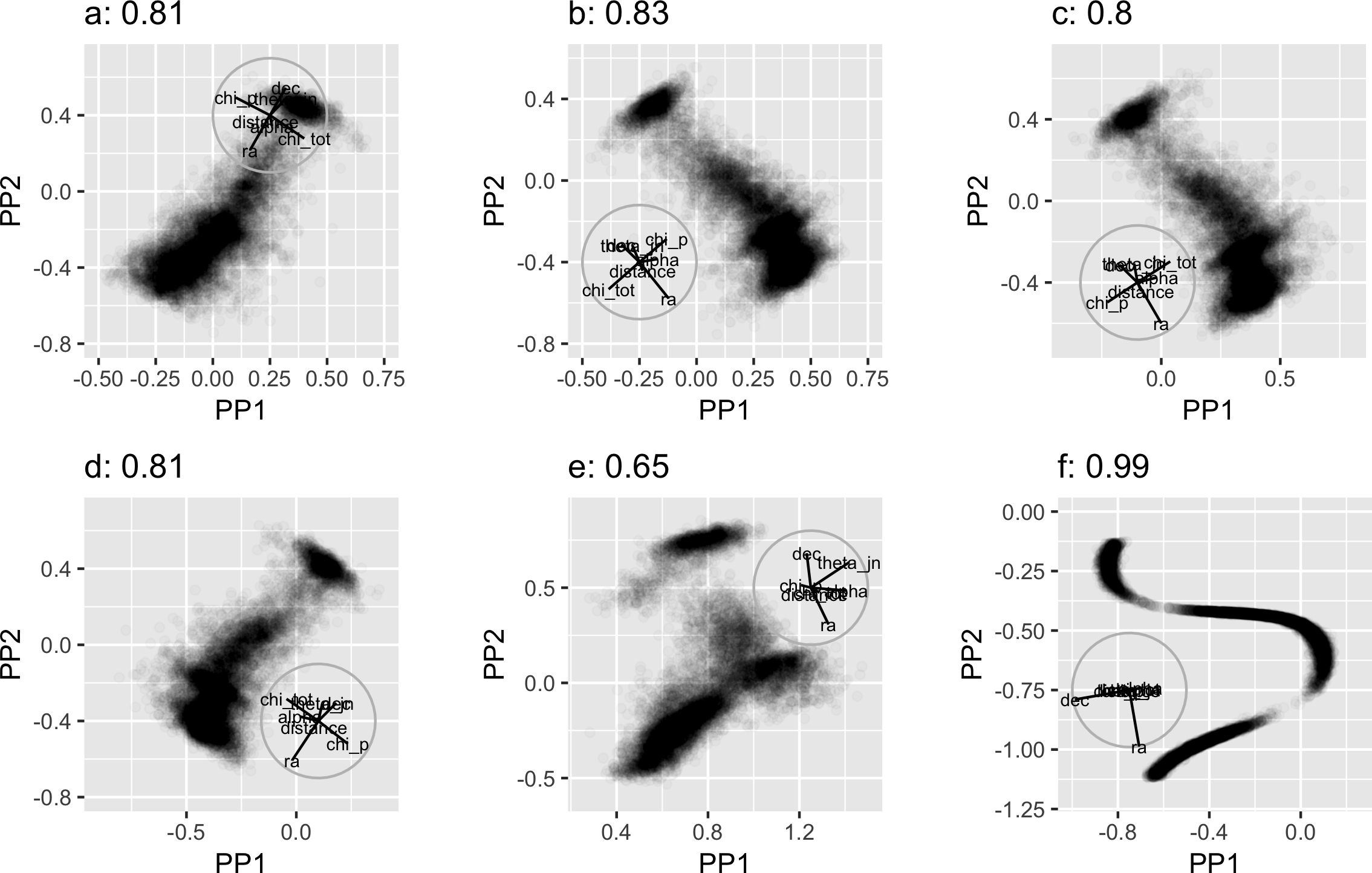} 

}

\caption{Final views identified in the dataset considering the seven dimensional parameter space (alpha, theta\_jn, chi\_tot, chi\_p, ra, dec, distance), differing only by randomly selected starting plane.}\label{fig:bbhGuided3}
\end{figure}

Figure \ref{fig:bbhGuided3} shows six maxima obtained by different
starts, for two types of parameters: first, spin related parameters
(i.e.~alpha, theta\_jn, chi\_tot and chi\_p), and second position
related parameters (i.e.~ra, dec and distance). Four of the six (a-d)
are almost identical, but not interesting projections. Projection f has
the highest PP index value but it is primarily the view seen in the
bivariate plot of dec and ra. While none of these projections are
particularly interesting on their own, moving between them can be
revealing.

\begin{figure}

{\centering \includegraphics[width=\textwidth]{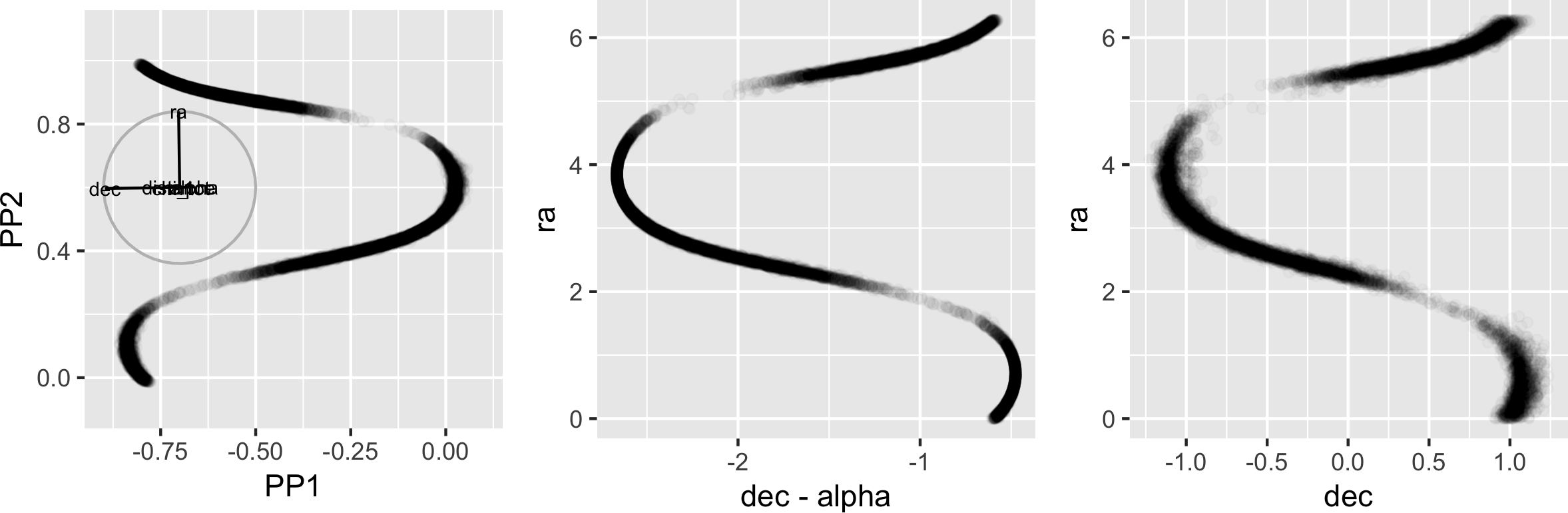} 

}

\caption{Manual reconstruction of an optimal projection (left), constructed by differencing alpha from dec in the original units against ra (middle), compared with the two main variables (right).}\label{fig:constructedExample}
\end{figure}

Choosing a different subset of variables reveals something new. The
subspace of m1, ra, chi\_tot, alpha, distance, dec produces a more
refined view of Figure \ref{fig:bbhGuided3} projection f.~When alpha
contributes in contrast to dec, the relationship between the points is
almost perfectly on a curve. This is shown in Figure
\ref{fig:constructedExample}. Manually reconstructing the optimal
projection (left plot) can be done by differencing the two parameters,
in their original units. This highlights the importance of improved
optimization, that would use tiny final steps to polish the view to a
finer optimal projection and possibly remove noise induced by small
contributions of many variables.

\hypertarget{instructions-for-applying-to-new-data}{%
\subsection{Instructions for applying to new
data}\label{instructions-for-applying-to-new-data}}

Applying these procedures to new datasets can be done using the guided
tour available in the \texttt{tourr} package. The typical steps required
are:

\begin{enumerate}
\def\labelenumi{\arabic{enumi}.}
\tightlist
\item
  Scale, standardize or sphere (principal components) the data.
\item
  Index function selection matching the type of structure that is
  interesting to detect. Any new function can be used as long as it
  takes a matrix as input and returns a single index value.
\item
  Call the guided tour with the data and index function:

  \begin{itemize}
  \tightlist
  \item
    for exploration this can be done via the \texttt{tourr::animate}
    function
  \item
    for recording the results \texttt{tourr::save\_history} should be
    used
  \end{itemize}
\item
  Explore how the results depend on choices made (index function,
  starting planes, optimization method, prior dimension reduction).
\end{enumerate}

These are steps followed in the above two applications and are
documented in the comments of the source code (U. Laa and Cook 2019a).
For simple usage examples, see documentation of the
\texttt{tourr::guided\_tour} function.

\hypertarget{discussion}{%
\section{Discussion}\label{discussion}}

The motivation for this work was to discover dependencies between
estimated parameters in multiple model fits in physics problems. This
paper shows how projection pursuit with the new indexes can help address
this problem. The results are encouraging, showing large potential for
discovering unanticipated relations between multiple variables.

All of the indexes fall short against some aspect of the ideal
properties of smoothness, squintability, flexibility, rotation
invariance and speed. The paper describes how these properties can be
assessed using tour methodology. Some potential fixes for the indexes
are discussed but there is scope for further developing the new indexes.
We recommend to use the spinebil (U. Laa and Cook 2019b) package when
developing new indexes. It includes the functionalities needed to
reproduce the assessments presented in this paper. While the current
focus is on two-dimensional index functions, indexes in the tourr
package apply to arbitrary projection dimension, and the methodology
introduced here could be applied to the assessment of index functions
where \(d>2\).

The work also reveals inadequacies in the tour optimization algorithm,
that may benefit from newly developed techniques and software tools.
Exploring this area would help improve the guided tours. As new
optimization techniques become available, adapting these to the guided
tour would extend the technique to a broader range of problems. The
current recommended approach is to first reduce the dimensionality, for
example by using PCA, taking care to preserve nonlinear structure, prior
to applying PP.

To apply the existing index functions in practice, we recommend to
either use the tourr package directly, or if interaction is required to
call the guided tour via the graphical interface available in the galahr
(U. Laa and Cook 2019c) package. This package supersedes the now
archived tourrGui (Huang, Cook, and Wickham 2012). Both packages contain
examples to show how the guided tour can be used with different index
functions.

\hypertarget{supplementary-material}{%
\section{Supplementary material}\label{supplementary-material}}

\begin{itemize}
\tightlist
\item
  This article was created with R Markdown (Xie, Allaire, and Grolemund
  2018), the code for the paper is available at (U. Laa and Cook 2019a).
\item
  Methods for testing new index functions as presented in this work are
  implemented in the R package spinebil (U. Laa and Cook 2019b).
\item
  The R package galahr (U. Laa and Cook 2019c) provides a graphical
  interface to the tourr package allowing for interactive exploration
  using the guided tour.
\item
  Additional explanations are available in the Appendix, covering
  details of

  \begin{itemize}
  \tightlist
  \item
    how the holes index was rescaled,
  \item
    estimations of the squint angle,
  \item
    a comparison of the computational performance of the index
    functions,
  \item
    testing the effect of index parameters on the results and
  \item
    suggestions how the index functions may be refined.
  \end{itemize}
\end{itemize}

\hypertarget{acknowledgments}{%
\section*{Acknowledgments}\label{acknowledgments}}
\addcontentsline{toc}{section}{Acknowledgments}

The authors gratefully acknowledge the support of the Australian
Research Council. We thank Rory Smith for help with the gravitational
wave examples, and German Valencia for useful comments. This article was
created with knitr (Xie 2015), R Markdown (Xie, Allaire, and Grolemund
2018) and bookdown (Xie 2016) with embedded code.

\hypertarget{references}{%
\section*{References}\label{references}}
\addcontentsline{toc}{section}{References}

\hypertarget{refs}{}
\leavevmode\hypertarget{ref-PhysRevLett119161101}{}%
Abbott, B. P., and others. 2017. ``GW170817: Observation of
Gravitational Waves from a Binary Neutron Star Inspiral.'' \emph{Phys.
Rev. Lett.} 119 (16): 161101.
\url{https://doi.org/10.1103/PhysRevLett.119.161101}.

\leavevmode\hypertarget{ref-Abbott2018exr}{}%
---------. 2018. ``GW170817: Measurements of neutron star radii and
equation of state.'' \url{http://arxiv.org/abs/1805.11581}.

\leavevmode\hypertarget{ref-AHC02}{}%
Ahn, J. S., H. Hofmann, and D. Cook. 2003. ``A Projection Pursuit Method
on the Multidimensional Squared Contingency Table.'' \emph{Computational
Statistics} 18 (3): 605--26.

\leavevmode\hypertarget{ref-minerva}{}%
Albanese, Davide, Michele Filosi, Roberto Visintainer, Samantha
Riccadonna, Giuseppe Jurman, and Cesare Furlanello. 2012. ``Minerva and
Minepy: A c Engine for the Mine Suite and Its R, Python and Matlab
Wrappers.'' \emph{Bioinformatics}, bts707.

\leavevmode\hypertarget{ref-As85}{}%
Asimov, D. 1985. ``The Grand Tour: A Tool for Viewing Multidimensional
Data.'' \emph{SIAM Journal of Scientific and Statistical Computing} 6
(1): 128--43.

\leavevmode\hypertarget{ref-BCAH05}{}%
Buja, A., D. Cook, D. Asimov, and C. Hurley. 2005. ``Computational
Methods for High-Dimensional Rotations in Data Visualization.'' In
\emph{Handbook of Statistics: Data Mining and Visualization}, edited by
C. R. Rao, E. J. Wegman, and J. L. Solka, 391--413.
http://www.elsevier.com: Elsevier/North Holland.

\leavevmode\hypertarget{ref-CBC92}{}%
Cook, D., A. Buja, and J. Cabrera. 1992. ``An Analysis of
Polynomial-Based Projection Pursuit.'' \emph{Computing Science and
Statistics} 24: 478--82.

\leavevmode\hypertarget{ref-CBCH94}{}%
Cook, D., A. Buja, J. Cabrera, and C. Hurley. 1995. ``Grand Tour and
Projection Pursuit.'' \emph{Journal of Computational and Graphical
Statistics} 4 (3): 155--72.

\leavevmode\hypertarget{ref-CBC93}{}%
Cook, Dianne, Andreas Buja, and Javier Cabrera. 1993. ``Projection
Pursuit Indexes Based on Orthonormal Function Expansions.''
\emph{Journal of Computational and Graphical Statistics} 2 (3): 225--50.
\url{http://www.jstor.org/stable/1390644}.

\leavevmode\hypertarget{ref-Cook:2018mvr}{}%
Cook, Dianne, Ursula Laa, and German Valencia. 2018. ``Dynamical
projections for the visualization of PDFSense data.'' \emph{Eur. Phys.
J.} C78 (9): 742. \url{https://doi.org/10.1140/epjc/s10052-018-6205-2}.

\leavevmode\hypertarget{ref-Cook:2007:IDG:1557227}{}%
Cook, Dianne, and Deborah F. Swayne. 2007. \emph{Interactive and Dynamic
Graphics for Data Analysis with R and Ggobi}. 1st ed. Springer
Publishing Company, Incorporated.

\leavevmode\hypertarget{ref-diaconis84}{}%
Diaconis, Persi, and David Freedman. 1984. ``Asymptotics of Graphical
Projection Pursuit.'' \emph{Ann. Statist.} 12 (3): 793--815.
\url{https://doi.org/10.1214/aos/1176346703}.

\leavevmode\hypertarget{ref-Eddy:1977}{}%
Eddy, William F. 1977. ``A New Convex Hull Algorithm for Planar Sets.''
\emph{ACM Trans. Math. Softw.} 3 (4): 398--403.
\url{https://doi.org/10.1145/355759.355766}.

\leavevmode\hypertarget{ref-1056714}{}%
Edelsbrunner, H., D. Kirkpatrick, and R. Seidel. 1983. ``On the Shape of
a Set of Points in the Plane.'' \emph{IEEE Transactions on Information
Theory} 29 (4): 551--59. \url{https://doi.org/10.1109/TIT.1983.1056714}.

\leavevmode\hypertarget{ref-FER13}{}%
Ferraty, F., A. Goia, E. Salinelli, and P. Vieu. 2013. ``Functional
Projection Pursuit Regression.'' \emph{Test} 22 (2): 293--320.
\url{https://search-proquest-com.ezproxy.lib.monash.edu.au/docview/1354331474?accountid=12528}.

\leavevmode\hypertarget{ref-corner}{}%
Foreman-Mackey, Daniel. 2016. ``Corner.py: Scatterplot Matrices in
Python.'' \emph{The Journal of Open Source Software} 24.
\url{https://doi.org/10.21105/joss.00024}.

\leavevmode\hypertarget{ref-f87}{}%
Friedman, J. H. 1987. ``Exploratory Projection Pursuit.'' \emph{Journal
of the American Statistical Association} 82 (1): 249--66.

\leavevmode\hypertarget{ref-FT74}{}%
Friedman, J. H., and J. W. Tukey. 1974. ``A Projection Pursuit Algorithm
for Exploratory Data Analysis.'' \emph{IEEE Transactionson Computers}
23: 881--89.

\leavevmode\hypertarget{ref-Grimm2016}{}%
Grimm, Katrin. 2016. ``Kennzahlenbasierte Grafikauswahl.'' Doctoral
thesis, Universität Augsburg.

\leavevmode\hypertarget{ref-mbgraphic}{}%
---------. 2017. \emph{Mbgraphic: Measure Based Graphic Selection}.
\url{https://CRAN.R-project.org/package=mbgraphic}.

\leavevmode\hypertarget{ref-hall89}{}%
Hall, Peter. 1989. ``On Polynomial-Based Projection Indices for
Exploratory Projection Pursuit.'' \emph{Ann. Statist.} 17 (2): 589--605.
\url{https://doi.org/10.1214/aos/1176347127}.

\leavevmode\hypertarget{ref-HWscagR}{}%
Hofmann, Heike, Lee Wilkinson, Hadley Wickham, Duncan Temple Lang, and
Anushka Anand. 2019. \emph{Binostics: Compute Scagnostics}.

\leavevmode\hypertarget{ref-CEM:CEM2568}{}%
Hou, Siyuan, and Peter D. Wentzell. 2014. ``Re-Centered Kurtosis as a
Projection Pursuit Index for Multivariate Data Analysis.'' \emph{Journal
of Chemometrics} 28 (5): 370--84.
\url{https://doi.org/10.1002/cem.2568}.

\leavevmode\hypertarget{ref-tourrGui}{}%
Huang, Bei, Dianne Cook, and Hadley Wickham. 2012. ``TourrGui: A
gWidgets Gui for the Tour to Explore High-Dimensional Data Using
Low-Dimensional Projections.'' \emph{Journal of Statistical Software} 49
(6): 1--12.

\leavevmode\hypertarget{ref-huber85}{}%
Huber, Peter J. 1985. ``Projection Pursuit.'' \emph{Ann. Statist.} 13
(2): 435--75. \url{https://doi.org/10.1214/aos/1176349519}.

\leavevmode\hypertarget{ref-JS87}{}%
Jones, M. C., and R. Sibson. 1987. ``What Is Projection Pursuit?''
\emph{J. Roy. Statist. Soc., Ser. A} 150: 1--36.

\leavevmode\hypertarget{ref-kr69}{}%
Kruskal, J. B. 1969. ``Toward a Practical Method Which Helps Uncover the
Structure of a Set of Observations by Finding the Line Transformation
Which Optimizes a New `Index of Condensation'.'' In \emph{Statistical
Computation}, edited by R. C. Milton and J. A. Nelder, 427--40. New
York: Academic Press.

\leavevmode\hypertarget{ref-2033241}{}%
Kruskal, Joseph B. 1956. ``On the Shortest Spanning Subtree of a Graph
and the Traveling Salesman Problem.'' \emph{Proceedings of the American
Mathematical Society} 7 (1): 48--50.
\url{http://www.jstor.org/stable/2033241}.

\leavevmode\hypertarget{ref-paperSRC}{}%
Laa, U., and D. Cook. 2019a.
\url{https://github.com/uschiLaa/paper-ppi}.

\leavevmode\hypertarget{ref-spinebil}{}%
---------. 2019b. \url{https://github.com/uschiLaa/spinebil}.

\leavevmode\hypertarget{ref-galahr}{}%
---------. 2019c. \url{https://github.com/uschiLaa/galahr}.

\leavevmode\hypertarget{ref-lckl2005}{}%
Lee, Eun-Kyung, Dianne Cook, Sigbert Klinke, and Thomas Lumley. 2005.
``Projection Pursuit for Exploratory Supervised Classification.''
\emph{Journal of Computational and Graphical Statistics} 14 (4):
831--46. \url{https://doi.org/10.1198/106186005X77702}.

\leavevmode\hypertarget{ref-ligoData}{}%
``LIGO.'' 2018. 2018.
\href{\%7Bhttps://dcc.ligo.org/public/0152/P1800115/005\%7D}{\{https://dcc.ligo.org/public/0152/P1800115/005\}}.

\leavevmode\hypertarget{ref-LOPERFIDO201842}{}%
Loperfido, Nicola. 2018. ``Skewness-Based Projection Pursuit: A
Computational Approach.'' \emph{Computational Statistics \& Data
Analysis} 120: 42--57.
\url{https://doi.org/https://doi.org/10.1016/j.csda.2017.11.001}.

\leavevmode\hypertarget{ref-Marsaglia68}{}%
Marsaglia, G. 1968. ``Random Numbers Fall Mainly in the Planes.''
\emph{Proceedings of the National Academy of Science}.

\leavevmode\hypertarget{ref-naito1997}{}%
Naito, Kanta. 1997. ``A Generalized Projection Pursuit Procedure and Its
Significance Level.'' \emph{Hiroshima Math. J.} 27 (3): 513--54.
\url{https://doi.org/10.32917/hmj/1206126967}.

\leavevmode\hypertarget{ref-PAN2000153}{}%
Pan, Jian-Xin, Wing-Kam Fung, and Kai-Tai Fang. 2000. ``Multiple Outlier
Detection in Multivariate Data Using Projection Pursuit Techniques.''
\emph{Journal of Statistical Planning and Inference} 83 (1): 153--67.
\url{https://doi.org/https://doi.org/10.1016/S0378-3758(99)00091-9}.

\leavevmode\hypertarget{ref-extracat}{}%
Pilhöfer, Alexander, and Antony Unwin. 2013. ``New Approaches in
Visualization of Categorical Data: R Package extracat.'' \emph{Journal
of Statistical Software} 53 (7): 1--25.
\url{http://www.jstatsoft.org/v53/i07/}.

\leavevmode\hypertarget{ref-posse95b}{}%
Posse, Christian. 1995a. ``Projection Pursuit Exploratory Data
Analysis.'' \emph{Computational Statistics \& Data Analysis} 20 (6):
669--87.
\url{https://doi.org/https://doi.org/10.1016/0167-9473(95)00002-8}.

\leavevmode\hypertarget{ref-posse95a}{}%
---------. 1995b. ``Tools for Two-Dimensional Exploratory Projection
Pursuit.'' \emph{Journal of Computational and Graphical Statistics} 4
(2): 83--100. \url{http://www.jstor.org/stable/1390759}.

\leavevmode\hypertarget{ref-rref}{}%
R Core Team. 2018. \emph{R: A Language and Environment for Statistical
Computing}. Vienna, Austria: R Foundation for Statistical Computing.
\url{https://www.R-project.org/}.

\leavevmode\hypertarget{ref-Reshef1518}{}%
Reshef, David N., Yakir A. Reshef, Hilary K. Finucane, Sharon R.
Grossman, Gilean McVean, Peter J. Turnbaugh, Eric S. Lander, Michael
Mitzenmacher, and Pardis C. Sabeti. 2011. ``Detecting Novel Associations
in Large Data Sets.'' \emph{Science} 334 (6062): 1518--24.
\url{https://doi.org/10.1126/science.1205438}.

\leavevmode\hypertarget{ref-JMLRv1715308}{}%
Reshef, Yakir A., David N. Reshef, Hilary K. Finucane, Pardis C. Sabeti,
and Michael Mitzenmacher. 2016. ``Measuring Dependence Powerfully and
Equitably.'' \emph{Journal of Machine Learning Research} 17 (212):
1--63. \url{http://jmlr.org/papers/v17/15-308.html}.

\leavevmode\hypertarget{ref-5508437}{}%
Rodriguez-Martinez, E., J. Yannis Goulermas, T. Mu, and J. F. Ralph.
2010. ``Automatic Induction of Projection Pursuit Indices.'' \emph{IEEE
Transactions on Neural Networks} 21 (8): 1281--95.
\url{https://doi.org/10.1109/TNN.2010.2051161}.

\leavevmode\hypertarget{ref-2014arXiv1401.7645S}{}%
Simon, Noah, and Robert Tibshirani. 2014. ``Comment on `Detecting Novel
Associations In Large Data Sets' by Reshef Et Al, Science Dec 16,
2011.'' \emph{arXiv E-Prints}, January, arXiv:1401.7645.
\url{http://arxiv.org/abs/1401.7645}.

\leavevmode\hypertarget{ref-Smith:2016qas}{}%
Smith, Rory, Scott E. Field, Kent Blackburn, Carl-Johan Haster, Michael
Pürrer, Vivien Raymond, and Patricia Schmidt. 2016. ``Fast and accurate
inference on gravitational waves from precessing compact binaries.''
\emph{Phys. Rev.} D94 (4): 044031.
\url{https://doi.org/10.1103/PhysRevD.94.044031}.

\leavevmode\hypertarget{ref-szekely2007}{}%
Székely, Gábor J., Maria L. Rizzo, and Nail K. Bakirov. 2007.
``Measuring and Testing Dependence by Correlation of Distances.''
\emph{Ann. Statist.} 35 (6): 2769--94.
\url{https://doi.org/10.1214/009053607000000505}.

\leavevmode\hypertarget{ref-barnett1981interpreting}{}%
Tukey, P.A., and J.W. Tukey. 1981. \emph{Graphical Display of Data in
Three and Higher Dimensions}. Wiley Series in Probability and
Mathematical Statistics: Applied Probability and Statistics. Wiley.
\url{https://books.google.com.au/books?id=WBzvAAAAMAAJ}.

\leavevmode\hypertarget{ref-wahba1990spline}{}%
Wahba, G. 1990. \emph{Spline Models for Observational Data}. CBMS-Nsf
Regional Conference Series in Applied Mathematics. Society for
Industrial; Applied Mathematics.
\url{https://books.google.com.au/books?id=ScRQJEETs0EC}.

\leavevmode\hypertarget{ref-tourr}{}%
Wickham, Hadley, Dianne Cook, Heike Hofmann, and Andreas Buja. 2011.
``tourr: An R Package for Exploring Multivariate Data with
Projections.'' \emph{Journal of Statistical Software} 40 (2): 1--18.
\url{http://www.jstatsoft.org/v40/i02/}.

\leavevmode\hypertarget{ref-scag}{}%
Wilkinson, L., A. Anand, and R. Grossman. 2005. ``Graph-Theoretic
Scagnostics.'' In \emph{IEEE Symposium on Information Visualization,
2005. INFOVIS 2005.}, 157--64.
\url{https://doi.org/10.1109/INFVIS.2005.1532142}.

\leavevmode\hypertarget{ref-WW08}{}%
Wilkinson, Leland, and Graham Wills. 2008. ``Scagnostics
Distributions.'' \emph{Journal of Computational and Graphical
Statistics} 17 (2): 473--91.
\url{https://doi.org/10.1198/106186008X320465}.

\leavevmode\hypertarget{ref-w16}{}%
Wood, S.N., N., Pya, and B. Säfken. 2016. ``Smoothing Parameter and
Model Selection for General Smooth Models (with Discussion).''
\emph{Journal of the American Statistical Association} 111: 1548--75.

\leavevmode\hypertarget{ref-knitr}{}%
Xie, Yihui. 2015. \emph{Dynamic Documents with R and Knitr}. 2nd ed.
Boca Raton, Florida: Chapman; Hall/CRC. \url{https://yihui.name/knitr/}.

\leavevmode\hypertarget{ref-bookdown}{}%
---------. 2016. \emph{Bookdown: Authoring Books and Technical Documents
with R Markdown}. Boca Raton, Florida: Chapman; Hall/CRC.
\url{https://github.com/rstudio/bookdown}.

\leavevmode\hypertarget{ref-rmarkdown}{}%
Xie, Yihui, J.J. Allaire, and Garrett Grolemund. 2018. \emph{R Markdown:
The Definitive Guide}. Boca Raton, Florida: Chapman; Hall/CRC.
\url{https://bookdown.org/yihui/rmarkdown}.

\bibliographystyle{spbasic}
\bibliography{bibliography.bib}

\end{document}